\documentclass[11pt]{article} 

\usepackage[utf8]{inputenc} 
\usepackage{booktabs,siunitx}
\usepackage{hyperref}
\usepackage{authblk}
\usepackage{natbib}
\usepackage{bbm}

\usepackage{geometry} 
\usepackage{graphicx} 
\usepackage{array}
\usepackage{xcolor}
\usepackage{multirow}
\usepackage{booktabs}

\usepackage{rotating}
\usepackage{dcolumn} 
\newcolumntype{d}{D{.}{.}{4}} 

\usepackage{booktabs} 
\usepackage{array} 
\usepackage{paralist} 
\usepackage{verbatim} 
\usepackage{subfig} 

\usepackage{amsmath}
\usepackage{amsthm} 
\usepackage{amssymb}
\usepackage{wasysym}
\usepackage{graphicx}
\usepackage{comment}
\usepackage{color}
\usepackage{multicol}
\usepackage{fancyhdr} 
\usepackage{sectsty}
\allsectionsfont{\sffamily\mdseries\upshape} 

\usepackage[nottoc,notlof,notlot]{tocbibind} 
\usepackage[titles,subfigure]{tocloft} 

\usepackage{authblk}

\title{Trend estimation and short-term forecasting of COVID-19 cases and deaths worldwide}

\author[a,1]{Ekaterina Krymova\footnote{coressponding author E-mail: ekaterina.krymova at epfl.ch}}
\author[a]{Benjamín B\'ejar} 
\author[b]{Dorina Thanou} 
\author[a]{Tao Sun}
\author[c]{Elisa Manetti}
\author[a]{Gavin Lee}
\author[c]{Kristen Namigai}
\author[a]{Christine Choirat}
\author[c]{Antoine Flahault}
\author[a]{Guillaume Obozinski}
 
\date{}

\affil[a]{Swiss Data Science Center, EPFL \& ETH Zürich, INN, Station 14, 1015 Lausanne, Switzerland} 
\affil[b]{Center for Intelligent Systems, EPFL, INF, Station 14, 1015 Lausanne, Switzerland} 
\affil[c]{Institute of Global Health, Faculty of Medicine, University of Geneva, 1202 Geneva, Switzerland}

\begin{document}
\definecolor{highlight}{HTML}{000000}
\definecolor{highlight1}{HTML}{000000}

\maketitle

\begin{abstract}
 Since the beginning of the COVID-19 pandemic, many dashboards have emerged as useful tools to monitor its evolution, inform the public, and assist governments in decision-making. Our goal is to develop a globally applicable method, integrated in a twice-daily updated dashboard that provides an estimate of the trend in the evolution of the number of cases and deaths from reported data of more than 200 countries and territories, as well as a seven-day forecast.  One of the significant difficulties to manage a quickly propagating epidemic is that the details of the dynamic needed to forecast its evolution are obscured by the delays in the identification of cases and deaths and by irregular reporting. Our forecasting methodology substantially relies on estimating the underlying trend in the observed time series using robust seasonal trend decomposition techniques. This allows us to obtain forecasts with simple, yet effective extrapolation methods in linear or log scale. We present the results of an assessment of our forecasting methodology and discuss its application to the production of global and regional risk maps.
\end{abstract}

\section*{Introduction}
It is of utmost importance for governments and decision-makers in charge of the healthcare system response to anticipate the evolution of the current COVID-19 pandemic \cite{velavan2020covid}. When accurate and reliable, predictions can be very informative to define appropriate policy measures and interventions, such as lockdown and containment measures, border closures, quarantines, school openings, and physical distancing. They are also useful to predict hospital surge capacity, in order to manage hospital resources \cite{lutz2019applying}.  Given that the pandemic is affected by these measures, testing policies, the appearance of new variants, the diffusions across borders, etc, long term forecasts are difficult and their usefulness remains unclear \cite{ioannidis2020forecasting}, whereas accurate short-term forecasts provide useful actionable information. Nevertheless, even short-term forecasts are far from trivial as recently evidenced by \cite{cramer2021evaluation}, where a simple baseline appears to be not so easy to beat on a one-week horizon.  

\begin{figure*}
\centering
\includegraphics[width=0.85\linewidth]{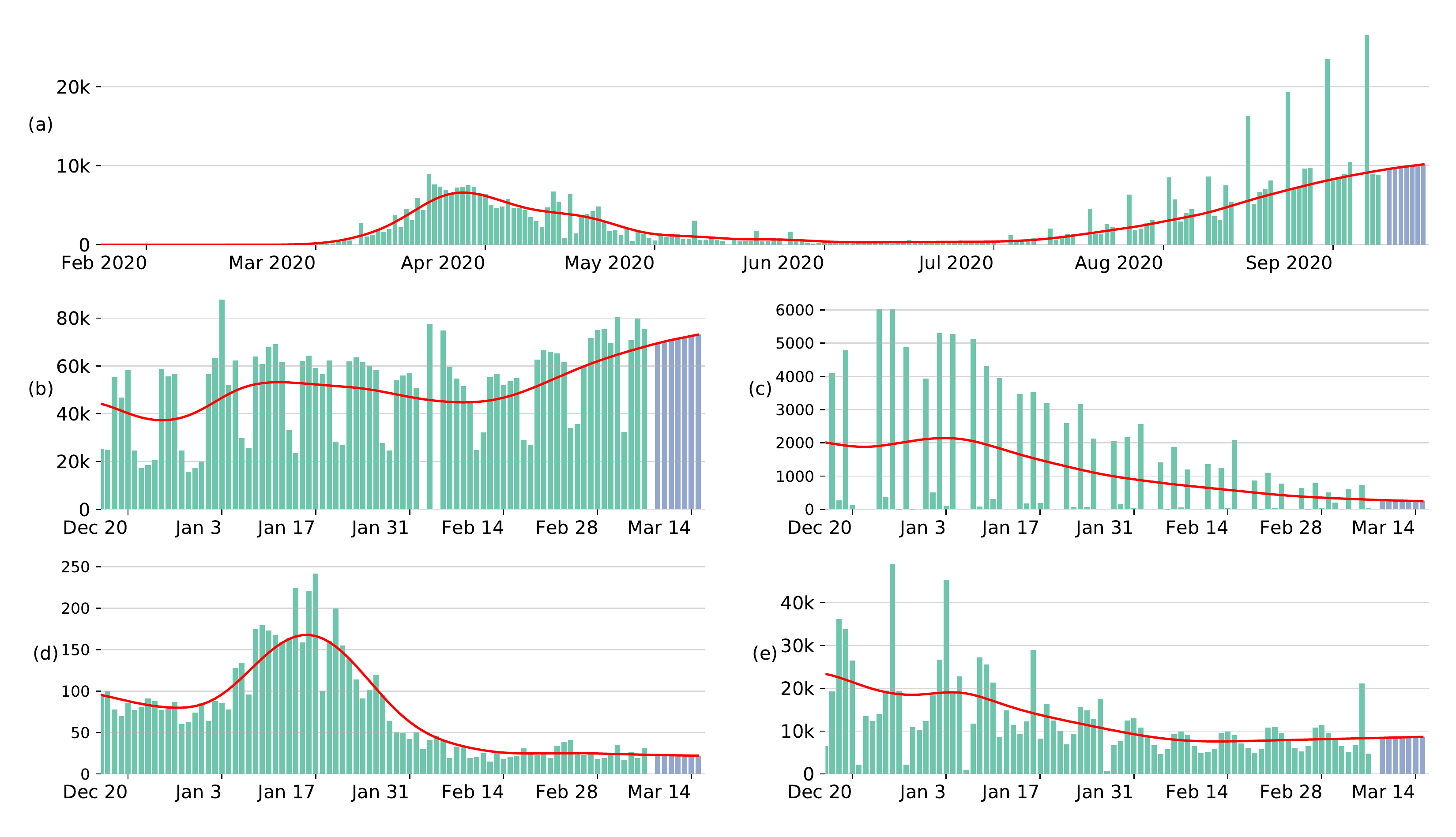}
\caption{Green bars correspond to daily cases. Blue bars show the forecast for the next 7 days. The red line shows the estimated trend smoothed together with the forecast. (a) JHU daily cases for Spain with forecasts starting from 10 September 2020 with a negative observation in June 2020 (not shown in the plot), visible outliers, and seasonality patterns in reporting starting from July. Daily cases in the last 3 months preceding 12-03-2021 for (b) Brazil, (c) Kansas (US), (d) China, (e) Germany. The observed number of cases (green), estimated trend (red), trend forecast for the following week (blue).}
\label{fig:trend}
\end{figure*}

In this work, we propose a general methodology to produce forecasts on a one-week horizon, which is applicable to close to 200 countries, and as many states/regions or provinces.  An additional challenge to achieve this goal is that the quality of the reported data varies significantly from country to country. This translates into different fluctuations and irregularities that can be observed in the reported time-series \cite{wilke2020predicting}.  Many countries do not report on a daily basis or delay their reports to particular days of the week. In particular, seasonal patterns with a weekly cycle are observed for many countries. \textcolor{highlight1}{In several countries, e.g.~in Switzerland, the number of reported cases shows a significant decline during or immediately after the weekend, which is probably due to the fact that, on those days, fewer patients get tested and/or the reporting is less active and thus delayed.} Also, it is important to note that, as illustrated in the data from Spain on Fig.~\ref{fig:trend} (a), seasonal patterns are non-stationary and can actually change in time, in particular, if the reporting policies change. Furthermore, delays in reporting, changes in death cause attribution protocols, as well as changes in testing policies lead to abrupt corrections that introduce backlogs on some days, such that a number of daily cases or deaths which are anomalously high or even negative are reported.  To take into account these  peculiarities, we propose a forecasting methodology that relies on estimating the underlying trend with a robust seasonal-trend decomposition method and using simple extrapolation techniques to make a forecast over a week.

\subsection*{Related work}

The problem of forecasting the evolution of the COVID-19 pandemic has attracted the attention of many researchers, institutions, and individuals across the globe. As a result, a significant number of dashboards have appeared that monitor and/or make predictions about the evolution of the pandemic based on past observations. All these efforts are also being leveraged to build ensemble predictive models for different regions in the World. For instance, the United States Center for Disease Control and Prevention  provides ensemble predictions for the US at the state level in  US Covid-19 Forecast Hub\footnote{ \url{viz.covid19forecasthub.org}}. Similarly, the German and Polish COVID-19 ForecastHub\footnote{\url{kitmetricslab.github.io}}  provides ensemble predictions at the regional level for Germany and Poland. And more recently, the European Centre for Disease Prevention and Control (ECDC) launched a European Covid-19 Forecast Hub\footnote{\url{covid19forecasthub.eu}} to provide short- and long-term ensemble forecasts for Europe. 
The considered modelling approaches rely on different data sources used (cases or/and death data, tests data, hospital data, mobility data etc.) and aim for forecasting horizons ranging from 1 week to months. Epidemiological compartmental models or models inspired by them are among the most popular ones for the forecasting task e.g., IHME \cite{covid2021modeling}, YYG-ParamSearch\footnote{\url{ covid19-projections.com}}, UMass-MechBayes \cite{gibson2020real}, \textcolor{highlight}{IEM\_Health-CovidProject \footnote{\url{iem-modeling.com}} and USC-SIkJalpha \cite{srivastava2020fast}}. Such models (e.g., SEIR \cite{allen2008mathematical}) split the population in different groups (age, demographics) and states (susceptible, infected, etc.) and model the transition dynamics of the population between the different states over time. The different parameters of the models can be deterministic or random, or be allowed to vary in time. Other approaches use statistical regression, e.g. UMich-RidgeTfReg\footnote{\url{gitlab.com/sabcorse/covid-19-collaboration}}, LANL \cite{castro2020coffee}, \textcolor{highlight}{curve fitting, e.g. RobertWalraven-ESG \footnote{\url{http://rwalraven.com/COVID19/}},} and machine learning, e.g. GT-DeepCOVID \cite{rodriguez2020deepcovid}, to learn a predictor from past observations, or time-series (e.g., ARIMA), e.g. \textcolor{highlight}{MUNI-ARIMA \footnote{\url{krausstat.shinyapps.io}}}, to learn a representation describing the evolution of the dynamics of the observed measurements, e.g.  CMU TimeSeries\footnote{\url{ github.com/cmu-delphi/covid-19-forecast}}, \cite{ahmad2021evaluating}. Some models make strong assumptions on the transmission dynamics \cite{castro2020turning} or specific assumptions on the effect of different policies, e.g. \cite{keskinocak2020impact,lemaitre2020scenario,bertozzi2020challenges}. 
These are just a few sample references and the list is by no means exhaustive. For a more complete list of recent related literature we refer the reader to
\cite{friedman2020predictive,cramer2021evaluation,bracher2020short, bracher2021national, ray2020ensemble}. 

Our approach differs from most of the other existing approaches in two ways.  First of all, given that the usefulness of long-term (e.g., several weeks or months) forecasts has been subject to debate, because of the complexity of the phenomenon and the impossibility of taking into account a number of important factors, we consider, like \cite{jewell2020predictive} “that short-term projections are the most that can be expected with reasonable accuracy”. We thus focus on the prediction of short-term (one to two weeks ahead) forecasts of daily numbers for deaths and cases. Second, instead of building directly a forecasting model, our approach implements first a trend estimation model from daily cases/deaths observations that make little or no assumptions about the underlying dynamics, and don't require to estimate a large number of parameters (as opposed to e.g., SEIR-type), hence are robust and easier to apply at a more global scale. Our forecast is then obtained from the trend estimated with a simple extrapolation scheme. Our trend estimates are of independent interest, and we  further use them to provide an independent estimate of the R-effective, which is an important measure for decision-making.

We apply the proposed algorithms to produce daily updated forecasts available on our public dashboard at \url{https://renkulab.shinyapps.io/COVID-19-Epidemic-Forecasting/}. \textcolor{highlight1}{Our dashboard,   together with an implemented forecasting methodology, has been in operation since the very beginning of the pandemic. The  forecasting methodology presented in the paper was put in production already in September 2020 and has therefore been producing results on the dashboard since.  The dashboard has been actively used by epidemiologists and global health experts to analyse the evolution of the epidemiological situation and to provide recommendations to several European governments. From the early phases of the pandemic, we delivered daily forecasts of cases and deaths for} 192 countries  as well as regions of several countries, including Switzerland, Canada, US, and France for which we also provide state or region level statistics and trend estimates. In addition, we provide estimates of \textcolor{highlight}{the prediction intervals based on the retrospective performance of the forecasting and } the effective reproduction number based on our trend estimate via the method of  \cite{huisman2020estimation} together with risk maps that assign to each country a color corresponding to its current epidemiological status (cf. Discussion).

\section*{Results}

\subsection*{Trend estimation}
To model a potentially quickly varying seasonal pattern and suppress the influence of outliers, we  implemented a piecewise trend estimation method based on the robust Seasonal Trend decomposition procedure based on LOESS (STL) \cite{robert1990stl}.  
STL is a filtering procedure for decomposing time series into trend, seasonal and residual components, which is furthermore robust to outliers. Specifically, the raw daily observations are modelled as: 
\begin{equation*} 
x_t = \tau_t+\delta_t+r_t 
\end{equation*}
where $\tau_t$  is a slowly changing trend, $\delta_t$ is a possibly slowly changing seasonal component and $r_t$ is a residual. Since the magnitude of the seasonal term can reasonably be expected to be proportional to the trend, allowing for the seasonal component to change with time is relevant here, especially since, as discussed, reporting patterns change in time in several time series. 
To obtain a more quickly adaptive algorithm, we use STL to produce separate trend estimates on windows of 6 weeks and recombine them. Outliers identified as a by-product of the STL procedure are removed, the corresponding counts are redistributed in recent history, and the trend estimation procedure is run one more time on the cleaned data (see the Methods Section for more details).

Fig.~\ref{fig:trend} illustrates the behavior of our trend estimation procedure for different countries that represent a certain diversity. In the cases of Germany and Brazil (b, e) the weekly seasonal effect is quite significant and clearly non-stationary, in particular between December 22nd and January 3rd, 2021. For China (d), no particular seasonal effect can be identified and several outliers seem to co-occur with the peak of the wave. The state of Kansas (US) (c) illustrates an example of a fairly irregular seasonal effect. In all these cases, the trend estimation proposed appears to be robust to outliers, to changes in seasonality or lack thereof, and adapts to the regularity of the underlying trend. Beyond a qualitative evaluation of the trend estimation, a quantitative evaluation is difficult for the lack of any ground truth, especially since the underlying dynamics of the trend in various countries and provinces are quite different. To some extent, the trend estimation proposed can be validated quantitatively via the forecasting algorithm which relies on it, since the quality of the forecast depends on the quality of the estimation of the trend.

\subsection*{Forecasting}

To predict cases and deaths one week ahead we propose to simply extrapolate linearly the daily trend, which was obtained with the above trend estimation algorithm (see Fig.~\ref{fig:trend}) either on the original or on the log-scale by preserving the most recent slope of the estimated trend. In the case of a decreasing trend slope the extrapolation is carried out in log-scale to prevent undershooting. For the case of an increasing trend the extrapolation is performed in linear scale to prevent overshooting. To forecast the number of deaths, some models have been using lagged cases as input. Given the diversity of situations in different territories, and the fact that the relation between deaths and cases was sometimes quite unclear or changing in a short amount of time, we used the same simple forecasting approach as for cases. See the SI for a discussion and references.

\begin{figure}
    \centering
    \includegraphics[width=0.7\textwidth]{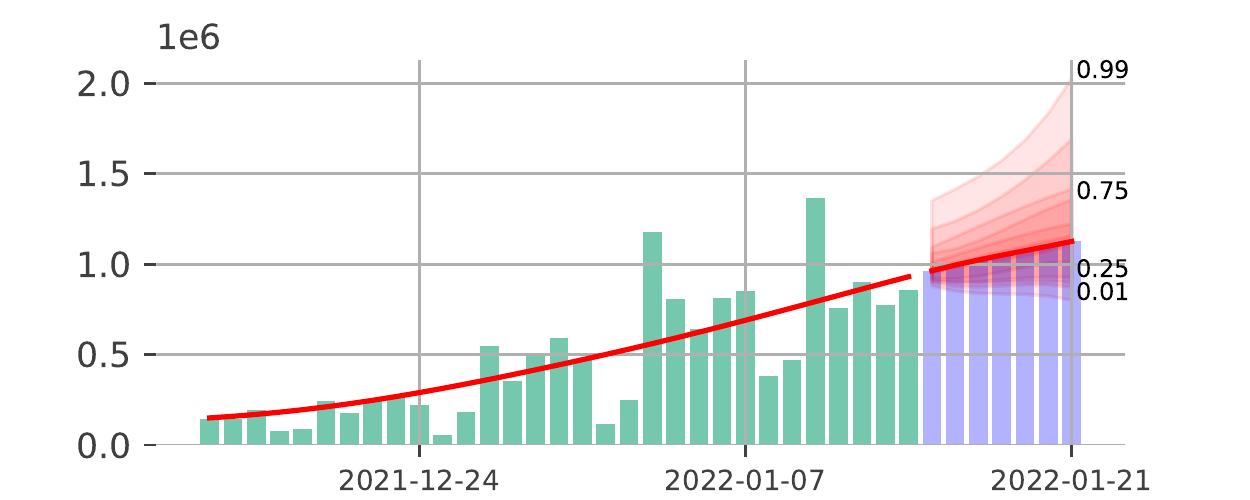}
    \caption{Illustration of the probabilistic forecast as a collection of nested intervals (red shaded regions) for the forecast of the number of cases in the US.}
    \label{fig:ci_us}
\end{figure}

\textcolor{highlight}{Following the recommendations of \cite{cramer2021evaluation} and the requests of different forecast hubs, we also produce a probabilistic forecasts for the weekly counts, under the form of a collection of 23 quantiles  corresponding to the levels $0.05 k$ for $k=1,\ldots,19$ and the extreme levels $\alpha=(0.001,0.025,0.975,0.99)$. These quantiles are estimated from quantiles estimates of appropriately normalized errors of our forecast on a recent history, that are extrapolated for extreme levels using a tail model. See Fig.~\ref{fig:ci_us} and Figs.~S12-14 in SI  for illustrations and  Materials and Methods for more detail.}

\subsection*{Evaluations}
\textcolor{highlight}{
We evaluate our forecasts of the number of new cases in two ways, first, by comparing them with the forecasts obtained by several methods submitted to the European Covid-19 Forecast Hub\footnote{\url{covid19forecasthub.eu}}, second, by comparing our forecast with a baseline on a larger set of countries, namely the naive forecast (used on several hubs) that assumes that the weekly number of cases remains constant over the following week. To obtain interval forecasts,  quantiles of the baseline predictive distribution  are estimated from symmetrized observed errors of the baseline as in the US and European Covid Forecast Hubs \cite{cramer2021evaluation}\footnote{\url{https://zoltardata.com/model/302}}. 
}

\textcolor{highlight}{
The comparison with the methods submitting of the European Covid-19 Forecast Hub is made in terms of MAE and average \emph{weighted interval score} (WIS) \cite{bracher2021evaluating} (see Materials and methods).} \textcolor{highlight1}{We also provide some details on the performance of our deaths forecasts in Section A of SI.}

To compare our method to the baseline we compute the relative improvement in mean absolute error (RMAE), relative improvement in median absolute error (RmedianAE), and relative improvement in average WIS (RWIS). The relative improvement is positive when the proposed forecasting method has a smaller error than the baseline. It can be thought of as a rate of decrease in error with respect to the baseline.

\subsubsection*{European countries, EU Hub}
\textcolor{highlight}{
In order to compare the performance of our method with other methods, we used the data available at the EU Covid Forecast Hub. The methods submitted to the Hub are aggregated in order to obtain a EuroCOVIDhub-ensemble method, and a baseline (EuroCOVIDhub-baseline) is available. The weekly forecasts can be submitted once a week.  Between April 1st and December 15th 2021, 43 submissions are available for both the EuroCOVIDhub-ensemble and the EuroCOVIDhub-baseline. We included in our comparison five other methods whose forecasts were all available for all the 31 countries included in the hub, on a common large subset of 32 weeks (i.e., 75\% of all weeks).   
These methods are: MUNI-ARIMA, IEM\_Health-CovidProject,  
USC-SIkJalpha \cite{srivastava2020fast}, RobertWalraven-ESG, 
ILM-EKF \footnote{\url{https://covid19forecasthub.eu/community.html}}. 
In order to obtain results comparable across different countries, we report the ratio of the MAE (or average WIS) of each method to the MAE (or average WIS) of the EuroCOVIDhub-baseline, following the reporting standards\footnote{\url{https://covid19forecasthub.eu/reports.html}} of the hub. (See SI Section A).   
}
\begin{figure}
    \centering
    \includegraphics[width=0.68\textwidth]{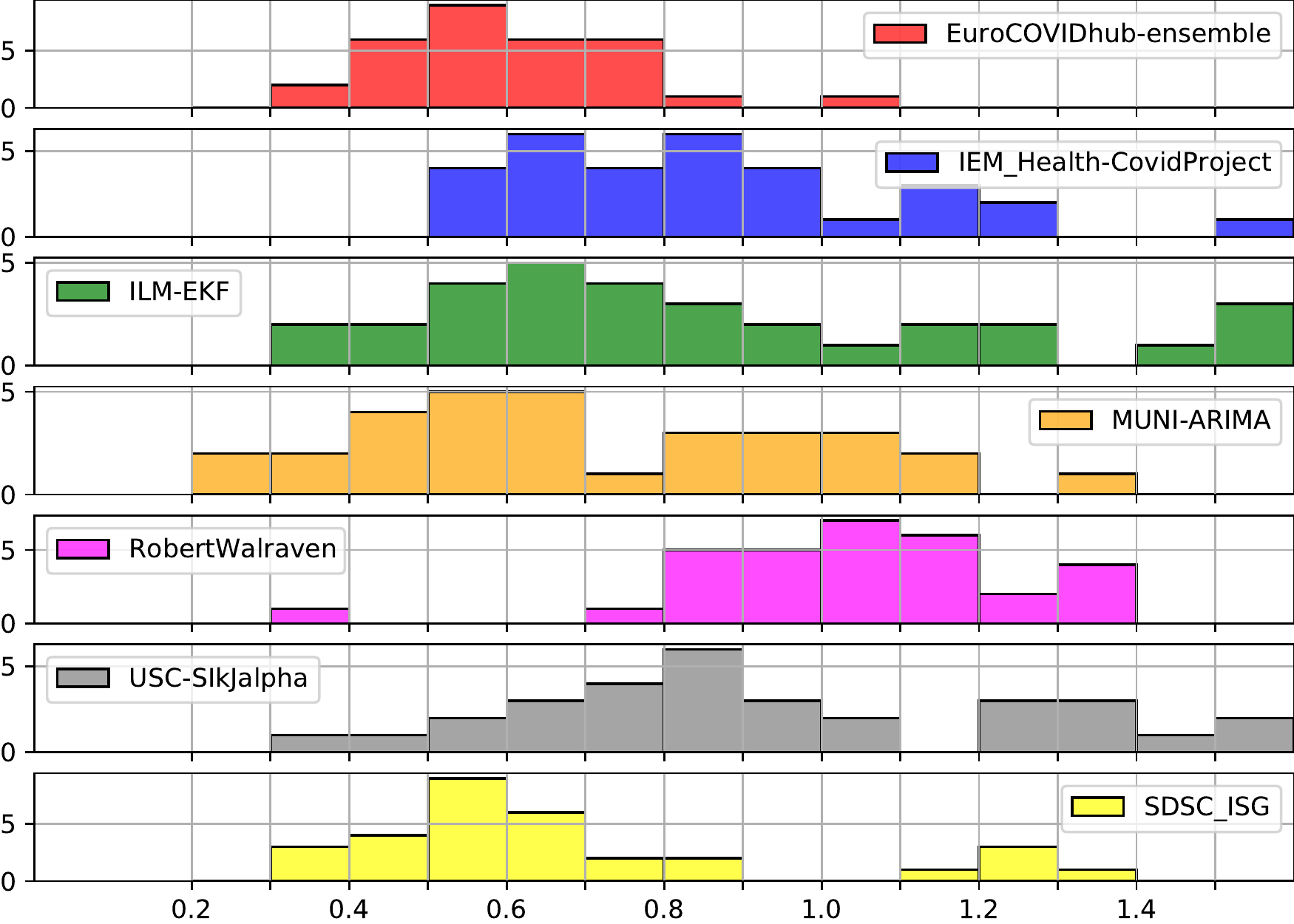}
    \caption{Histograms for the average WIS (in x-axis) of  1 week ahead forecasts for the 31 EU Hub countries.}
    \label{fig:wis_eu}
\end{figure}

\textcolor{highlight}{
 The results show that the proposed method (SDSC\_ISG) performs well for most of the countries, e.g. see the histogram of average WIS in Fig.~\ref{fig:wis_eu} (for MAE see Fig.~S1 in SI). 
 It is clear that in terms of average WIS, the performance of our method is one of the best ones and is close to the performance of the ensemble method and MUNI-ARIMA.}
 
 \textcolor{highlight}{
 Additionally, we ranked the methods according to their performance from 1 to 7, where 1 corresponds to the method with the smallest value of MAE or average WIS. The results can be summarized as follows:
for average WIS  {our method outperforms all other methods (including the ensemble) for 8 countries, it ranks second or best for 16 of them, and it is among the top three for 25 countries}; for MAE  {it ranks first for 9 countries, second or best for 18, and within the top three for 25}.  More details can be found in SI Section A and Tables S1-S2. 
Additionally, we used our methodology to perform 2 weeks ahead forecasts and obtained similar results (see Tables S3-S4 and Fig.~S2-S3 in SI). Given that our forecast is based on a simple extrapolation of our trend estimate, this suggests that the trend estimate is accurate even on the boundary of the period where data is available. 
}

\subsubsection*{Global comparison with a baseline}

For countries that report new cases with irregular delays, it is difficult to know whether the discrepancy between the forecast and reported weekly numbers is due to errors of the forecast or the fact that the reported numbers actually do not reflect accurately the current number of new cases. 

We, therefore, present the main evaluation of our forecasting strategy on a restricted set of 80 countries, which report sufficiently frequently with a relatively low number of outliers. 
These countries were selected based on a set of criteria that are  independent from our trend estimation and forecast methodology (see Materials and Methods). Nevertheless, we present the evaluation on the full list of countries in SI Section A. We use the data provided by Johns Hopkins University \cite{dong2020interactive} after April 1st, 2020, which corresponds approximately to the date after which all countries started reporting regularly. 

\begin{figure*}
\centering
\includegraphics[width=.98\linewidth]{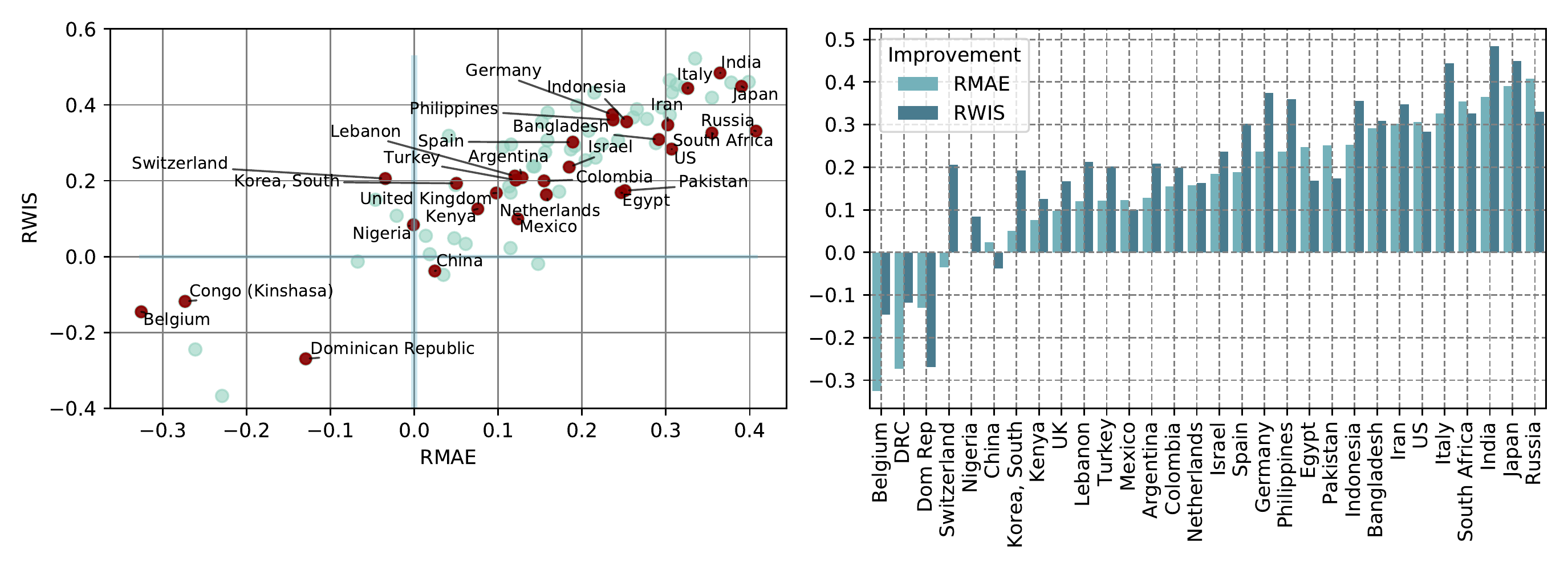}
\caption{(left) Scatter plot of RMAE and RWIS on one week ahead forecast for the selected subset of 80 countries, with points in red corresponding to the 30 countries with either larger populations or larger population density  included in the bar plot on the right.}
\label{fig:rmae}
\end{figure*}

\textcolor{highlight}{
For the 80 selected countries, we performed a retrospective analysis from April 1st, 2020 till December 15, 2022. For each day in this period, we forecast the total number of cases over the week following that day using our methodology and the baseline using the data that was available at that date (and thus without corrections made a posteriori). As ground-truth we used the weekly data available on January 10, 2022.  We evaluate our forecast by reporting, for each country  the RMAE,  the RmedianAE, the relative improvement in coverage, and RWIS.
The detailed  evaluation results for the full list of considered countries can be found in SI Section A. In Fig.~\ref{fig:rmae} (left) we display a scatter plot of the relative improvement in terms of RMAE and RWIS of our proposed forecast methodology over the baseline for the subset of 80 regularly reporting countries and for one week ahead forecast. As it can be seen, our method outperforms the baseline in both metrics for most of the selected countries. The same RMAE and RWIS values are displayed in Fig.~\ref{fig:rmae} (right) for 30 countries with either large populations or large population density, where the impact of the pandemic is potentially more important in terms of scale. }

\textcolor{highlight}{
  Out of the 80 countries, 72 (i.e., 90\%) show an improvement in MAE, 66 (82.5\%) show an improvement in median AE, 71 (88.75\%) show an improvement in WIS, and 68 countries show an improvement in both MAE and WIS. Only 5 countries do not show an improvement in either of these criteria. We also measure the coverage of the estimated prediction intervals, where our forecast is more accurate than the baseline for 66 countries. There are 53 out of 80 countries for which our method shows an improvement in all four metrics (MAE, median AE, average WIS, and coverage). This is the case, for example, for the US where the improvement in MAE and WIS over the baseline is 25\%.} We note that in the work of  \cite{cramer2021evaluation}, which compares forecasting algorithms focusing on US data, only 6 algorithms out of 23 achieve an improvement of more than 20\% in MAE over the baseline (for forecasts on horizons of 1 to 4 weeks; see Table 2 in that paper). 

The countries, for which our method did not perform better than the baseline, typically have long plateaus that the baseline benefits from, or/and have quickly changing seasonality patterns and direction of the trend, where the simple and robust baseline makes smaller errors. We analysed how the AE varies as a function of the growth rate of the trend and  evaluated to what extent, as soon as the trend is not flat, our method produces improved forecasts compared to the baseline 
 \textcolor{highlight}{(see Section F and Fig.~S4 in SI). Our method outperforms the baseline predictor when the growth rate is larger than 3\% in absolute value, which shows that the proposed forecast is informative as soon as the trend is not flat.}

\section*{Discussion}
\textcolor{highlight}{ 
The comparisons with the forecasts submitted to the European Covid Forecast Hub and the baseline demonstrate that the proposed forecasting methodology performs well.
It should be noted that, as a forecasting method, the considered baseline is uninformative in the sense that it does not attempt to characterize the evolution of the curve. In spite of this, as reported in \cite{cramer2021evaluation}, it is not easy to outperform this type of baseline in terms of pure predictive accuracy. Thus, any forecasting method characterizing the evolution of the curves which improves over this baseline can be useful. 
}
Our forecasts are available in the US Covid-19 Forecast Hub, European Covid-19 Forecast Hub, and German and Polish COVID-19 Forecast Hub\footnote{\url{https://kitmetricslab.github.io/forecasthub/forecast}}.  

Apart from producing forecasts and estimates of prediction intervals, the trend estimates that we obtain are of independent interest. We use them in particular to produce a stable estimate of the R-effective (R-eff for short). The R-eff measures the expected number of people that can be infected by an individual at any given time \cite{mahase2020covid} and has been used as a key indicator in this pandemic. Since its estimation requires essentially to solve a deconvolution problem \cite{huisman2020estimation}, it is quite sensitive to the irregularities in the data. In the original paper the authors use LOESS smoothing in order to decrease their influence. In our case, we propose to apply the deconvolution based on our piecewise robust STL trend estimate.

\begin{figure}
\centering\includegraphics[width=1\linewidth]{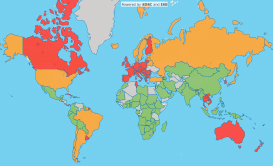}
\caption{World risk map from 2022-03-22.}
\label{fig:maps}
\end{figure}

Finally, we use our estimates of the R-eff together with our forecasts to produce global daily risk maps according to the following scheme: If the number of tests as reported in Our World in Data is available and is above 10,000 per 1M individuals, we compare  the  prediction for the number of weekly cases per 100K  and R-eff with corresponding thresholds to color code the map. Green is assigned  if the number of weekly cases per 100K inhabitants was below 30, orange is assigned if it is above 30 and the epidemic curve is descending (R-eff<0.9), and red is assigned if it is above 30 the epidemic curve is ascending or plateauing (R-eff>0.9). Other organisations use similar thresholds for the cumulative rate per 100K. For instance, our choice of a threshold coincides with the upper value of the 3rd level (out of 7) in the ECDC map of the geographic distribution of Covid cases. The value of the R-eff is not taken into account in the risk assessment (color code) of a country when the incidence numbers are low since its estimation becomes less reliable. Besides, it theoretically converges to 1 at the end of an epidemic, and in such regime it is no longer an indicator of the severity of the pandemic. As a consequence, if no test data is available, or the number of tests are below 10,000 per 1M population, the region is colored in grey, meaning that the data is missing/unreliable and no risk assessment can be made. An example of a risk map is given in Fig.~\ref{fig:maps}. These maps are useful to compare the levels of epidemic activity between countries based on a discrete color code. These comparisons can be insightful, especially if the R-eff and different non-pharmaceutical interventions of countries are taken into account. 

The fact that our model  makes minimal assumptions about the data is an advantage to make it applicable to a large number of countries and regions. But there are of course downsides. In particular, our models only take into account a fairly limited amount of information, and in particular no indicators of mobility, prophylactic measures, or lockdowns are taken into account. Our models can however detect the effect of changes of behavior, which can be quite informative for decision-makers. For example, when Ireland faced a second wave and the decision of lockdown was made on the October 21st, 2020 our models predicted exponential growth for the 7 following days. However, almost the day after the curve broke and on  October 25th the R-eff was clearly below 1, with a descending epidemic trend. That could not be attributed to the lockdown measures only four days after being taken. 

\section*{Conclusion}

We have proposed a methodology for trend estimation and short-term forecasts of the evolution of the number of Covid-19 cases and deaths, which is broadly applicable to a large number of different countries, states, and regions. Beyond its use to produce forecasts, our trend estimation method is of independent value, as it aims at providing a clear view of the current local evolution of the trend. Estimating the recent behavior of the trend is important as a tool to assess the current epidemic situation and to be able subsequently to analyze the effect of various measures. We use in particular our trend estimate to produce our own estimates of the R-effective curves, which are in turn used to produce risk maps. 
For the forecast, our evaluation shows that, \textcolor{highlight}{ 1) the methodology performs well compared to several methods submitted to the European Covid Forecast Hub, 2)} for the 80 selected countries in which we can reliably use weekly data as ground truth, we outperform the baseline in a large fraction of countries.

\section*{Materials and methods} 
\subsection*{Preprocessing}
Before applying the trend estimation model to the data, we remove negative values corresponding to reassessment of the counts, while making sure that the cumulative counts are preserved by appropriately scaling our estimates. We infer which zero counts on a given day correspond to missing reports, and we eventually impute the counts corresponding to these missing reports. The corresponding procedures are detailed in SI Section B.  
\subsection*{STL}
Our trend estimation algorithm leverages the Seasonal-Trend decomposition procedure based on LOESS (STL) proposed by \cite{robert1990stl}. STL consists of an iterative procedure that alternates between the estimation of the seasonal and trend components, each of them being estimated using a LOESS model, as well as the reestimation of importance weights associated with each observation for robust estimation. More precisely, the algorithm consists of two nested loops: The inner loop comprises several steps involving moving average and LOESS nonparametric regressions \cite{cleveland1988locally} in order to estimate the seasonal and trend components for the current set of robustness weights. The importance weights are then updated in the outer loop based on the residuals after the update of the seasonal and trend components. The procedure is repeated for a number of iterations.

\subsection*{Trend estimator} First, we smooth with STL separately all intervals of 6 weeks  that are starting every three weeks from the end of the time series (so that each interval is half overlapping with the previous one). In the absence of outliers (which should be  detected by the robust STL procedure), each trend estimate computed on a six weeks interval is rescaled to account for the same total number of counts as the observed data, and to obtain the final trend estimate, on each disjoint three weeks interval defined from the end of the time-series, we compute a pointwise weighted combination of the two overlapping trend estimates computed on this interval. In the presence of outliers, these are first identified by our method and redistributed in the past. After that, the same procedure as explained before is applied to the corrected data. For further details, we refer to SI Section C.
 
 \subsection*{Probabilistic forecast}
\textcolor{highlight}{
We produce probabilisic forecasts under the form of a collection of 23 quantiles for the predicted average daily counts, which produce as well a collection of nested prediction intervals of the form $[q_{\alpha},q_{1-\alpha}]$.
These quantiles are estimated based on empirical quantiles of the retrospective deviations of the daily forecast $f_{h,t}$ from the actual weekly rolling mean  ${\bar{x}}_{t+h}=\frac{1}{7}\sum_{k=-3}^3 x_{t+h+k}$  for horizons $h=1,\dots,H$ and 19 levels $\alpha=0.05,0.10,\dots,0.95$, normalized by $\sqrt{f_{t,h}}$, namely the scaled errors $$({\bar{x}}_{t+h}-f_{t,h})/{\sqrt{f_{t,h}}}.$$ 
The length of the history for estimation of the 19 quantiles  was set to $40+H$ days. The normalization by $\sqrt{f_{t,h}},$ which performed better than $f_{t,h}$ or $1$, is motivated by a Poisson error distribution model. For the lowest/highest levels ($\alpha=0.01,0.025,0.975,0.99$), we extrapolated the quantiles based on an exponential tail model whose scaling parameter is estimated from the other estimated 19 quantiles.
Finally, we shift all quantiles $\tilde{q}_{\alpha_i,h}$ on the scaled errors by a constant to enforce that $\tilde{q}_{0.5,h}=0.$ This is motivated by the fact that we expect our forecast to be close to the conditional best median forecast. 
The prediction quantiles are finally 
of the form $q_{\alpha_i,h}=f_{t,h}+\tilde{q}_{\alpha_i,h} \sqrt{f_{t,h}},$ where $f_{t,h}$ is the current forecast. See Fig.~\ref{fig:ci_us} and Figs.~S12-14 in SI for illustrations.\\
Similarly, for the weekly total number of cases/deaths $k$ weeks ahead, $k=\{1,2\},$ the ground truth can be computed as $X_{t+4+7(k-1)}=\sum_{h=1}^{7}{x}_{t+h+7(k-1)}$ and the point forecast are just $F_{t,4+7(k-1)}=\sum_{h=1}^{7} f_{h,t+7(k-1)}$. The probabilistic forecast can be computed using the same approach as above by replacing $\bar{x}_{t+h}$ by $X_{t+4+7(k-1)}$ and $f_{t,h}$ by $F_{t,4+7(k-1)}.$ In that case, the horizon is in weeks instead of days.
}
\subsection*{Evaluation metrics for point forecasts}
\textcolor{highlight}{
If $\bar{x}_t=\frac{1}{7}\sum_{k=-3}^3 x_{t+h+k}$  is as before the rolling mean of the number of daily new cases over a week and $f_t$ is the corresponding point forecast (which we identify with the median forecast), the absolute error of $f_t$ is ${\rm AE}(f_t) = |f_t-\bar{x}_t|$. We consider, as evaluation metrics, the mean absolute error (MAE) and the median absolute error over the evaluation period. Given a  baseline  $b_t=\bar{x}_{t-7}$, the relative MAE is defined as 
$${\rm RMAE}= ({\rm MAE}(b)-{\rm MAE}(f))/{\rm MAE}(b).$$ 
Similarly, one can define RmedianAE. 
}
\subsection*{Evaluation metrics for probabilistic forecasts}
 
Following the methodology presented in \cite{cramer2021evaluation}, we evaluate our probabilistic forecasts using proper scoring rules defined for forecasts taking the form of a collection of quantiles or equivalently of nested intervals, namely the \emph{weighted interval score} (WIS).\\
 The \emph{interval score} \cite{bracher2021evaluating} at level $\alpha\in(0,1)$ for the interval $[\ell,u]$ and observation $\xi$ is defined as 
  \begin{align*}
    IS_{\alpha}([\ell,u],\xi) &= u-\ell+ \frac{2}{\alpha}\big[ (\ell-\xi)\mathbbm{1}\{\xi<\ell\} + (\xi-u)\mathbbm{1}\{\xi>u\}\big],
  \end{align*} 
where $\mathbbm{1}\{\cdot\}$ is $1$ if the condition is satisfied and zero otherwise.
The \emph{weighted interval score} (WIS) \cite{bracher2021evaluating} is a proper scoring rule for probabilistic forecast, which is defined as follows: for a number of levels $A=\{\alpha_1,\dots,\alpha_K\}$, $\alpha_i \in [0,0.5)$ and the corresponding estimated quantiles of the predictive distribution $P$, defined as $q_{\alpha} = \inf \{q \mid P(\Xi\leq q)\geq \alpha)$ for the level $\alpha_i\in \{\alpha_1,\dots,\alpha_K,0.5,1-\alpha_K,\dots, 1-\alpha_1\}$, where $\Xi$ is the random variable associated with the observation $\xi,$ as follows:
\begin{equation*}
    {\rm WIS}(P,A,\xi) = |\xi-q_{0.5}| + \sum_{k=1}^{K}\alpha_k IS_{2\alpha_k}([q_k,q_{2K+2-k}],\xi).
\end{equation*}
The average WIS is defined as the mean of the WIS for the predictive quantiles of distributions $P_t$ constructed to predict $\bar{x}_t$ over the times $t$ in the estimation interval, i.e.,
${\rm MWIS}=\frac{1}{T} \sum_{t=1}^T {\rm WIS}(P_t,A,\bar{x}_t).$
Relative improvement in MWIS (RWIS) is defined in similarly as RMAE.

\subsection*{Selection of countries with more reliable data}
For our main results, we kept 80 countries whose reports of cases are sufficiently frequent, have only few missing values and a limited number of outliers. We proceeded as follows. First, we excluded 52 countries that reported cases on less than 70\% of the days, since these countries have either a very small number of cases or are reporting very irregularly, and among the remaining countries, the 39 countries for which more than five consecutive days were missing. Then, we performed robust outlier detection (described in SI Section D) to estimate the number of outliers in each time series and we excluded the 20 countries with the largest number of outliers among the remaining ones. It is important to note that the selection criteria proposed here are independent of our trend estimation and forecast methodology.

\section*{Code and Data availability}
\textcolor{highlight}{Code and evaluations are accessible at {\small{\url{https://github.com/ekkrym/CovidTrendModel}}}.}
We use the publicly available data collected by Johns Hopkins University, which consists of countrywise daily cases and deaths. Regional-level data for Canada, Switzerland, and France are obtained from the JHU repository, the Specialized Unit for Open Government Data of the Canton of Zürich, and the French National Health Agency, respectively. 

\section{Acknowledgments*}{The development of the dashboard was partly funded by the Fondation Privée des Hôpitaux Universitaires de Genève. We would like to thank Fernando Perez-Cruz for several discussions and constructive comments which led to methodological improvements.}
 
 
\bibliographystyle{unsrtnat}
\bibliography{covid}

\appendix
\section*{Supplemental Information}
\addcontentsline{toc}{section}{Appendices}
\renewcommand{\thesubsection}{\Alph{subsection}}
\setcounter{figure}{0} 
\setcounter{table}{0} 

\renewcommand\thetable{S\arabic{figure}} 
\renewcommand\thefigure{S\arabic{figure}} 
\section*{A. Evaluation}

\subsection*{Global comparison with the baseline}
 
\begin{figure}
\centering
\includegraphics[width=0.90\textwidth]{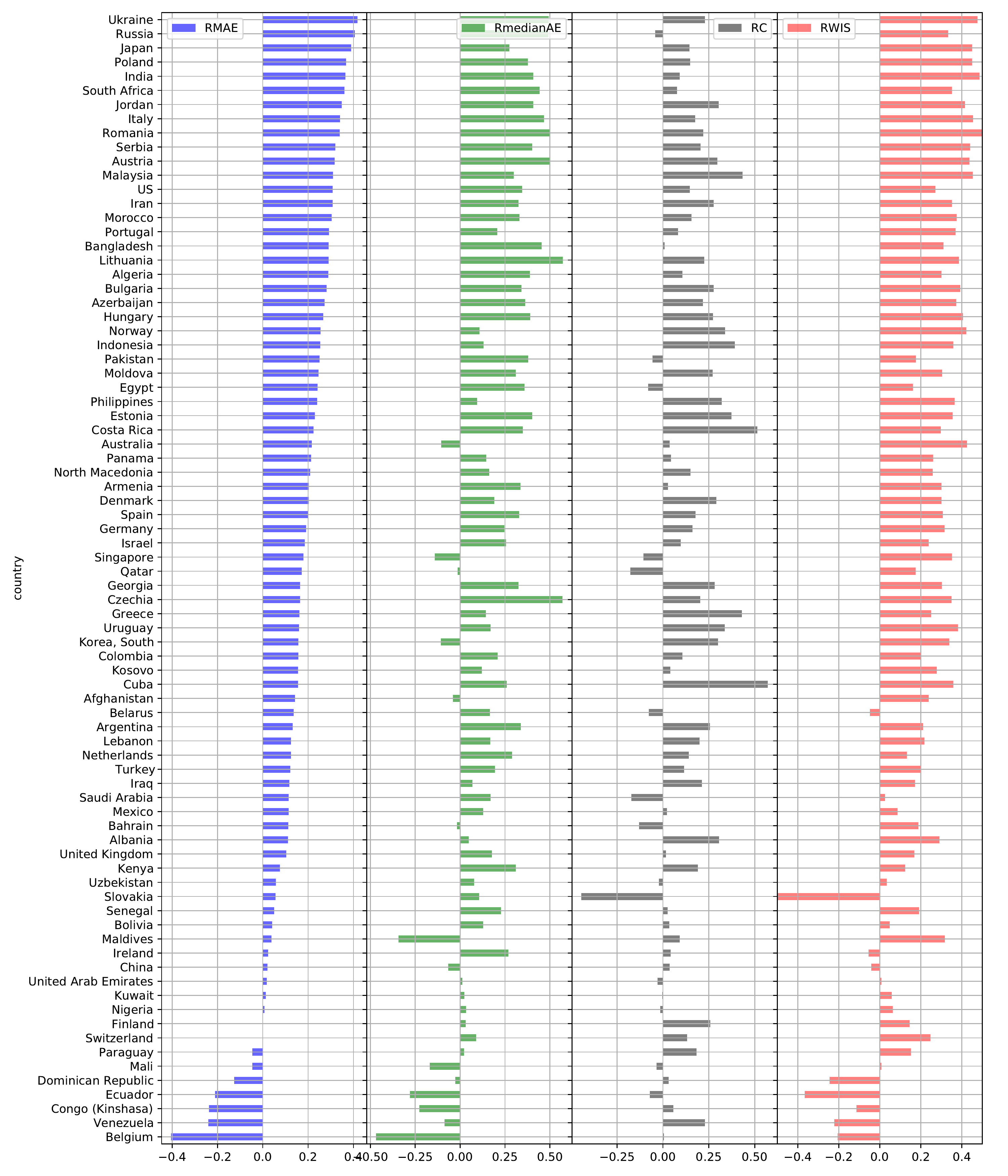}
\caption{RMAE: For 90 \% of countries of the selected set our method outperforms the baseline in MAE for average weekly cases. (b) RmedianAE: For 82.5\% of countries of the selected set our method outperforms the baseline in median AE for average weekly cases. (c) R0: For 80\% of the countries of the selected set, our method performs better than the baseline in total coverage  (d) RWIS;  For 88.75\% of countries of the selected set our method outperforms the baseline in average WIS for average weekly cases}
\label{fig:rmae_80}
\end{figure}

\begin{figure}
\centering
\includegraphics[width=0.95\textwidth]{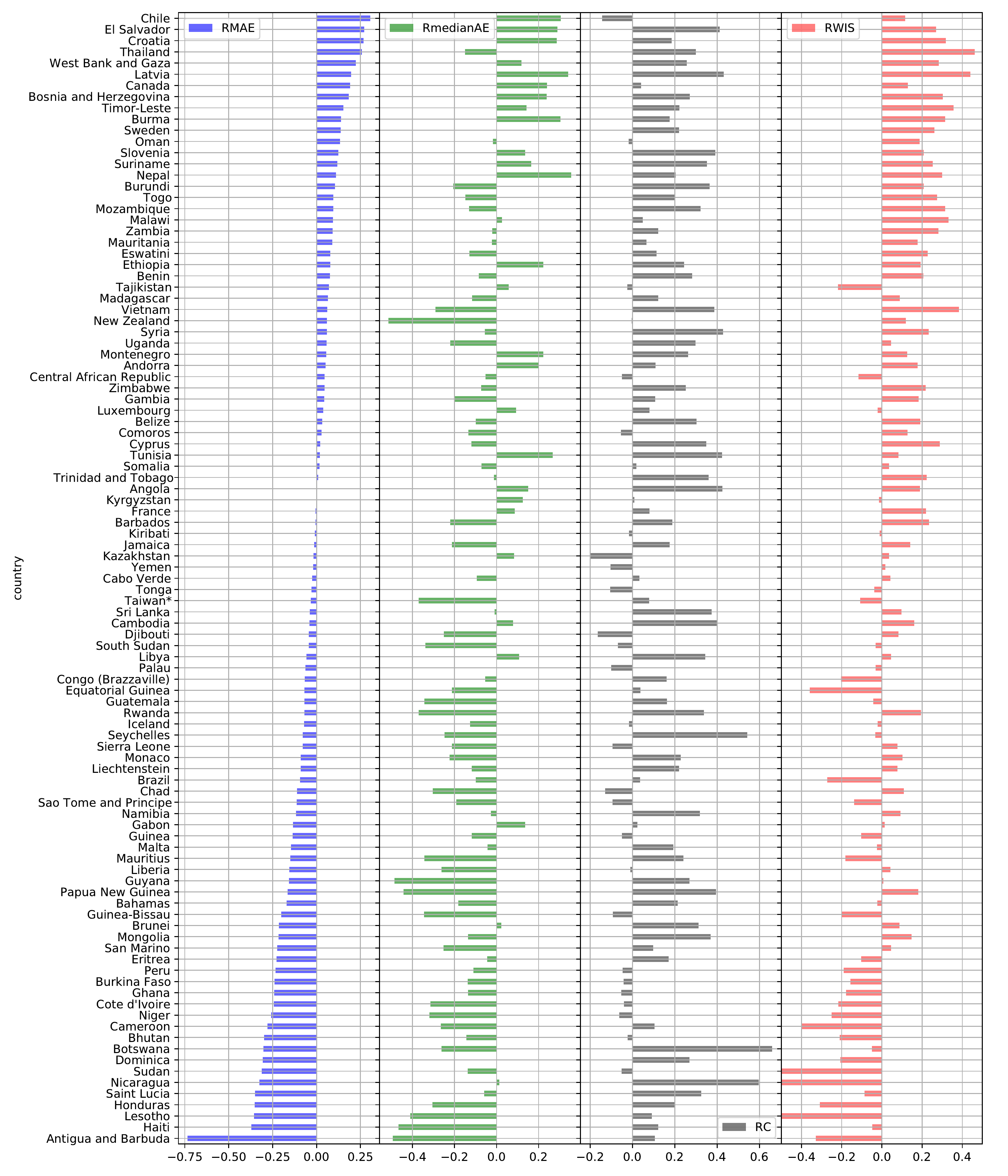}
\caption{ Evaluation scores for the 101 countries, corresponding to the countries not included in the list of 80 countries considered in Section that have more than 10\% of non-zero daily observations (this excludes countries with more than 90\% of zero daily observations: Marshall Islands, Grenada, Vanuatu, Tanzania, Fiji, Saint Kitts and Nevis, Micronesia, Samoa, the Holy See, Solomon Islands, Laos, Saint Vincent and the Grenadines) (a)  RMAE. (b) RmedianAE. (c) RC (improvement in total coverage). (d) RWIS.
}
\label{fig:rmae_rest}
\end{figure}

We evaluate our method by reporting, for each country the RMAE, the RmedianAE, the relative improvement in mean total coverage (see Section E), and average WIS. The relative improvement in mean total coverage is computed as  ${\rm RC}= ({\rm MC}(s)-{\rm MC}(b))/{\rm MC}(b)$ from the coverages  {$MC$} of the methods, since one aims for higher coverage by the  estimates of the confidence intervals.\\
In Fig.~\ref{fig:rmae_80},  we illustrate the evaluation for the 80 countries with reliable data. In Fig.~\ref{fig:rmae_rest}, we illustrate the evaluation score for the 101 countries that did not pass our evaluation criteria. The evaluation on the remaining countries (with the exclusion of 11 countries which had a particularly low number of cases) shows that there remains 42 countries, for which our method still obtains a better MAE than the baseline, only 29 for which we improve in median AE, 63 for which RWIS is improved and 74 with improved total coverage. Many of the countries for which our method performs clearly worse than the baseline are in fact countries with a fairly low number of cases, which were not the focus of our modelling efforts. This is justified by the fact that accurate forecasts are not critical for these countries as long as their number remains low. Furthermore, the corresponding time series of a significant number of these countries have numerous irregularities of the reporting and backlogs which makes it harder to associate the target average weekly number with the true underlying trend. In that case, the simple baseline forecast, which is the average value of the previous week, appears to be closer to the target than the forecast obtained after trend estimation.  Different smoothing techniques would be needed to produce better trend estimation for these countries, which take into account the discrete nature of the count and their Poissonian distribution. 

\subsection*{Comparison with the forecasts submitted to European Covid Forecast Hub, cases}
 
\begin{figure}
    \centering
    \includegraphics[width=0.7\textwidth]{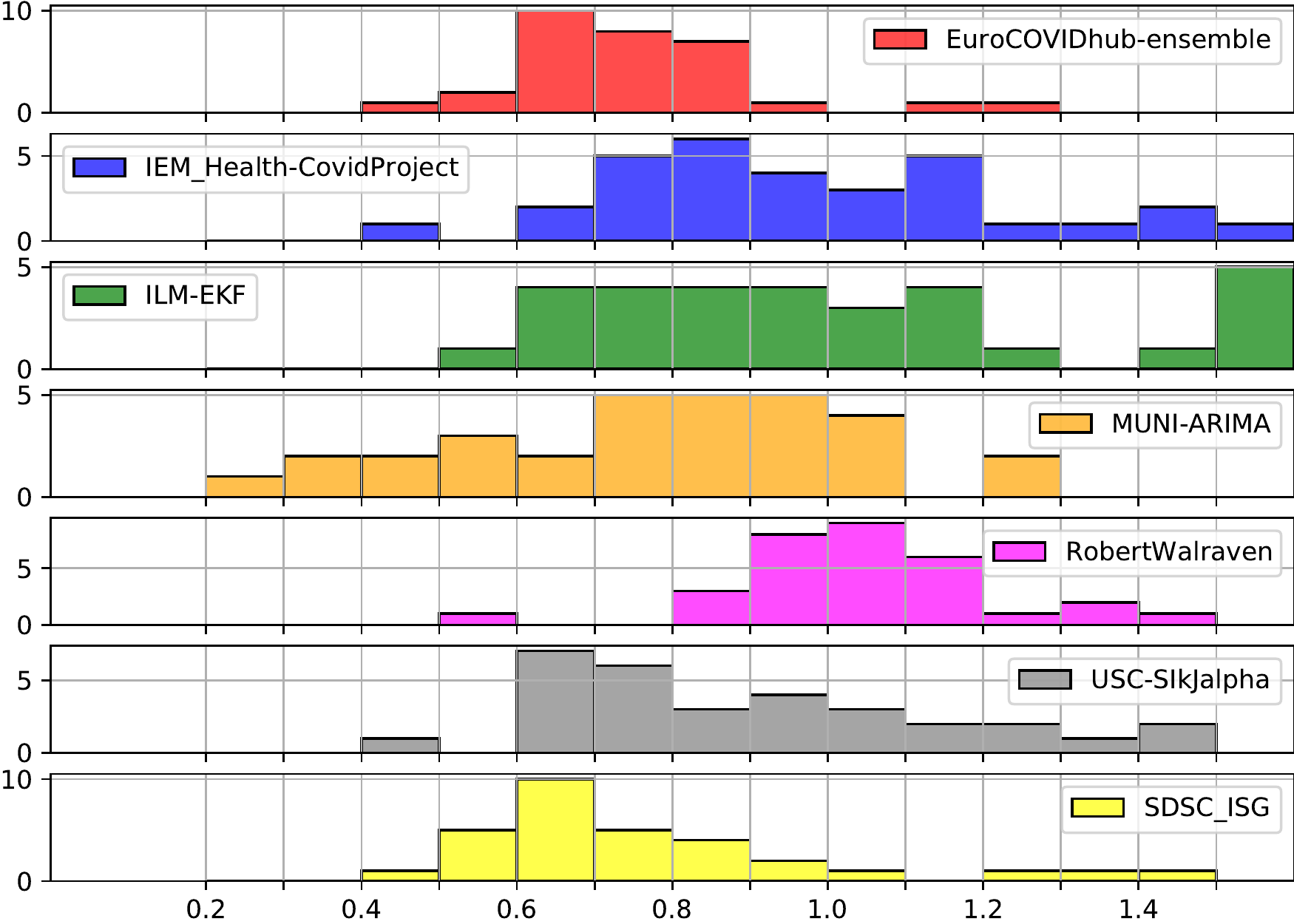}
    \caption{Histograms for the MAE (in x-axis) based on  1 week ahead cases forecasts for 31 European countries}
    \label{fig:mae_1w}
\end{figure}

\begin{figure}
    \centering
    \includegraphics[width=0.7\textwidth]{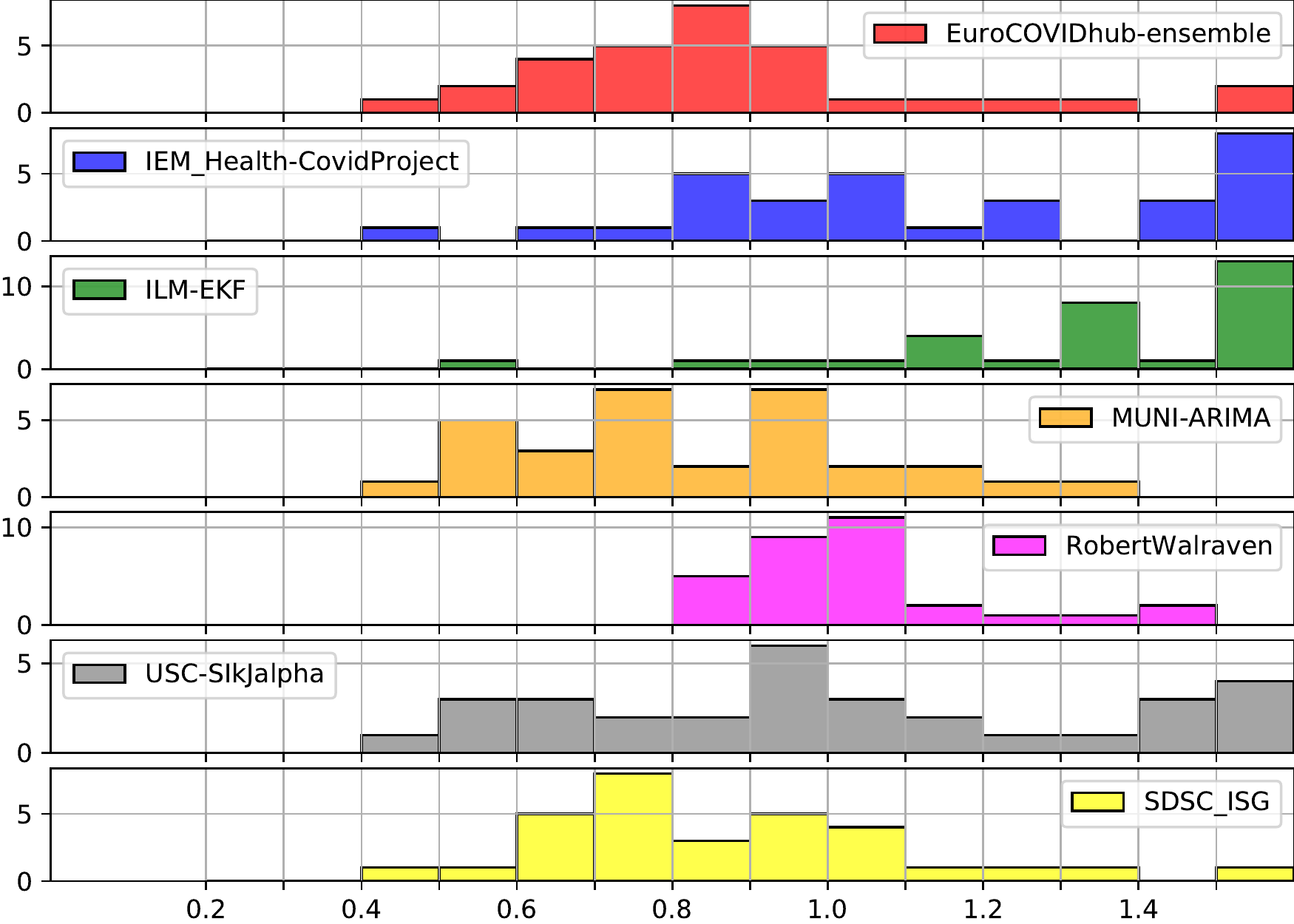}
    \caption{Histograms for the MAE (in x-axis) based on 2 week ahead cases forecasts for 31 European countries}
    \label{fig:mae_2w}
\end{figure}

\begin{figure}
    \centering
    \includegraphics[width=0.7\textwidth]{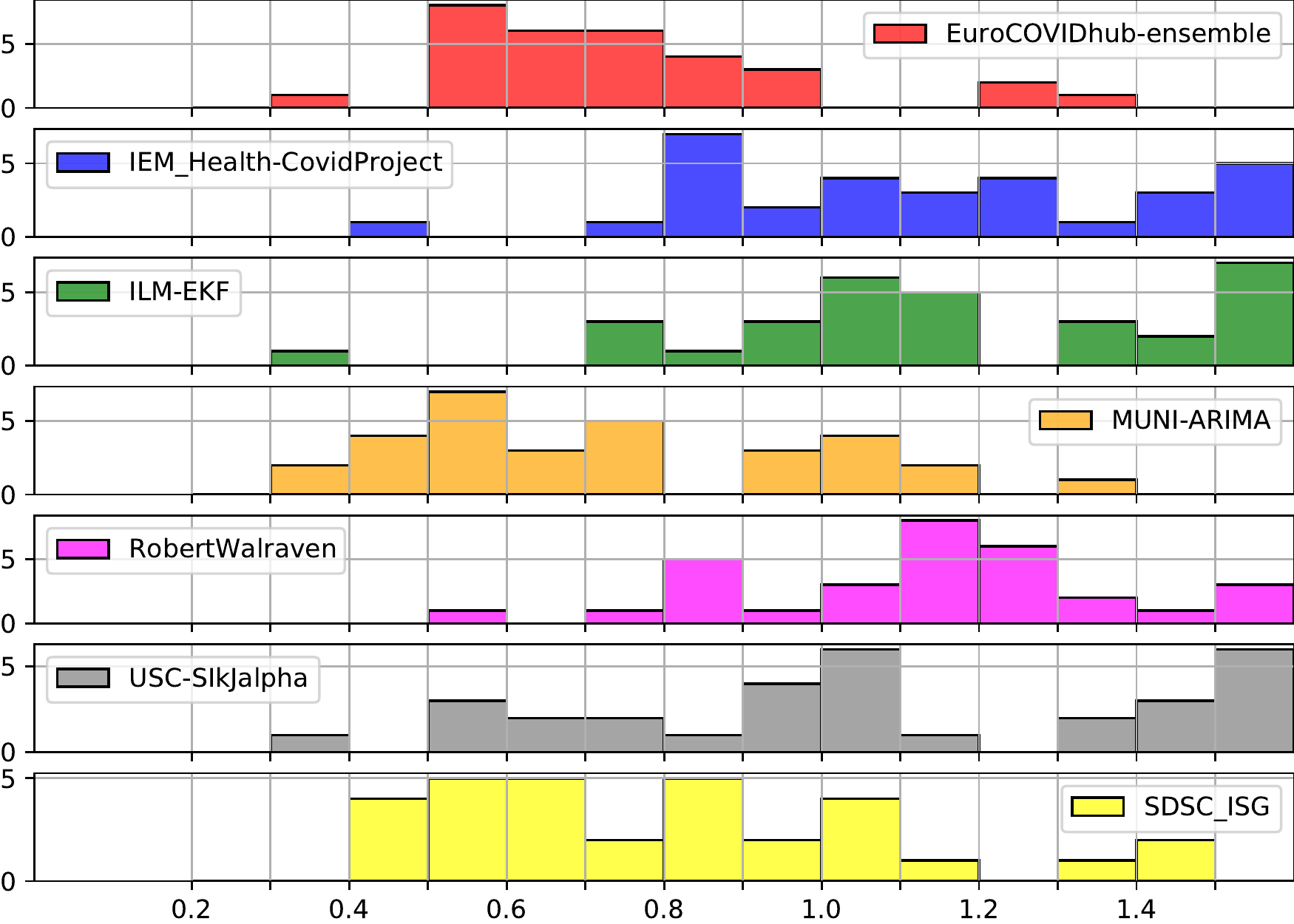}
    \caption{Histograms for the average WIS (in x-axis) based on 2 week ahead cases forecasts  for 31 European countries}
    \label{fig:wis_2w}
\end{figure}

The list of methods includes (with abbreviations used in the Tables \ref{tab:AE_w1}-\ref{tab:WIS_w2} below):
\begin{itemize}
    \item EuroCOVIDhub-ensemble (EUHub-ens,\\ \url{https://covid19forecasthub.eu/visualisation.html}), 
    \item EuroCOVIDhub-baseline (\url{https://covid19forecasthub.eu/visualisation.html}), 
    \item MUNI-ARIMA (MUNI, \url{https://krausstat.shinyapps.io/covid19global/}), 
    \item IEM\_Health-CovidProject (IEM\_Health, \url{https://iem-modeling.com/}), 
    \item USC-SIkJalpha (USC, \url{https://scc-usc.github.io/ReCOVER-COVID-19/#/}), 
    \item RobertWalraven-ESG (RW, \url{http://rwalraven.com/COVID19/Model}),  
    \item ILM-EKF (ILM, \url{https://github.com/Stochastik-TU-Ilmenau}),
    \item the proposed method (SDSC\_ISG, \\ \url{https://renkulab.shinyapps.io/COVID-19-Epidemic-Forecasting/}). 
\end{itemize}
Fig. \ref{fig:mae_1w}-\ref{fig:wis_2w} show histograms of the errors of the methods with respect to the baseline with $0.1$-wide bins. For visualization purposes, errors greater than 1.55  were set to 1.55 and therefore contribute to the last bin.
One week ahead forecasts MAE measured in multiples of the baseline MAE and average WIS measured in multiples of the baseline WIS are presented in Tables \ref{tab:AE_w1} and \ref{tab:WIS_w1}.
Two week ahead forecasts MAE measured in multiples of the baseline MAE and average WIS measured in multiples of the baseline WIS are presented in Tables \ref{tab:AE_w2} and \ref{tab:WIS_w2}.

\subsection*{Death forecasting: motivation and comparison with European Covid Forecast Hub submissions}
\begin{figure}
\centering
\includegraphics[width=1.1\textwidth]{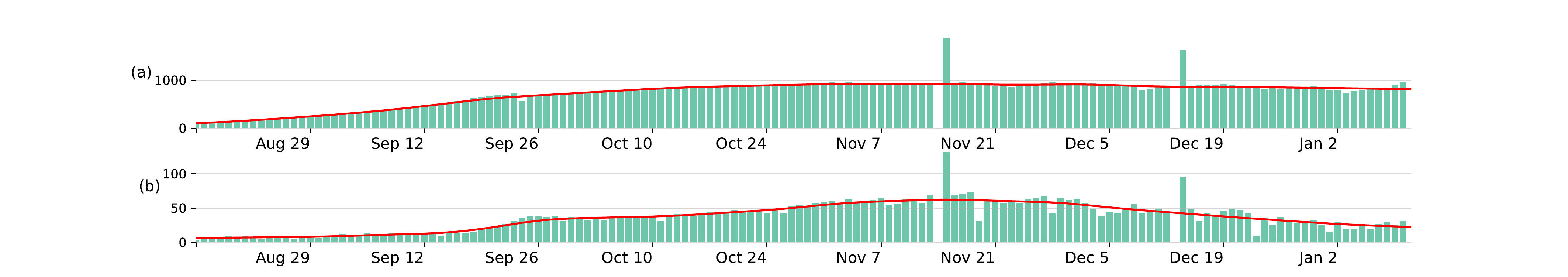} 
\caption{Cases in (a) and deaths numbers in (b) in Egypt in the end of 2021: the growth and decrease of death numbers is not preceeded by the similar behavior in the cases.}
\label{fig:egypt}
\end{figure}

\textcolor{highlight1}{ 
At first sight, in a very simple theoretical model, the number of deaths should be related to the number of cases, and correspond simply to the fraction of the cases that did not survive. The strategy of estimating the deaths from cases has been particularly successful for the USA at the country and state levels. Among the models participating in the US COVID Forecast Hub, \texttt{epiforecasts-ensemble1} contains a model which estimates deaths from a convolution of cases, the model \texttt{MIT\_Crit\_Data-GBCF} takes 3-weeks lagged deaths and cases numbers as a part of the input, etc. Forecasting the number of deaths from a lagged case curve was one of our first approaches, but the diversity of situations encountered across the world and in time for a particular region makes it that this strategy fails in a number of cases. The relation between lagged cases and the number of deaths is sometimes quite unclear: for example, if we consider the evolution of the number of cases and deaths in Egypt in November-December 2021 that we show in Fig.~\ref{fig:egypt}, the number of cases is almost not changing while the number of death is increasing and then decreasing. There are many reasons why the relation between the number of cases might be more complicated, be non-stationary and potentially change relatively quickly: as the virus circulates it affects different groups in the population who are more or less fragile and who protect their senior more or less well, the testing and reporting policies of some countries have sometimes changed quite quickly (including reporting policies for deaths), there is an effect of the vaccination (which is however on  a sufficiently slow timescale that it can be reestimated over time), there is the emergence of new variants, etc. Taking into account all of the above, we use the same strategy for the number of deaths forecasting as for cases, i.e. we estimate the trend based solely on the previous deaths observations and predict future numbers by the simple extrapolation.}

\begin{figure}
    \centering
    \includegraphics[width=0.7\textwidth]{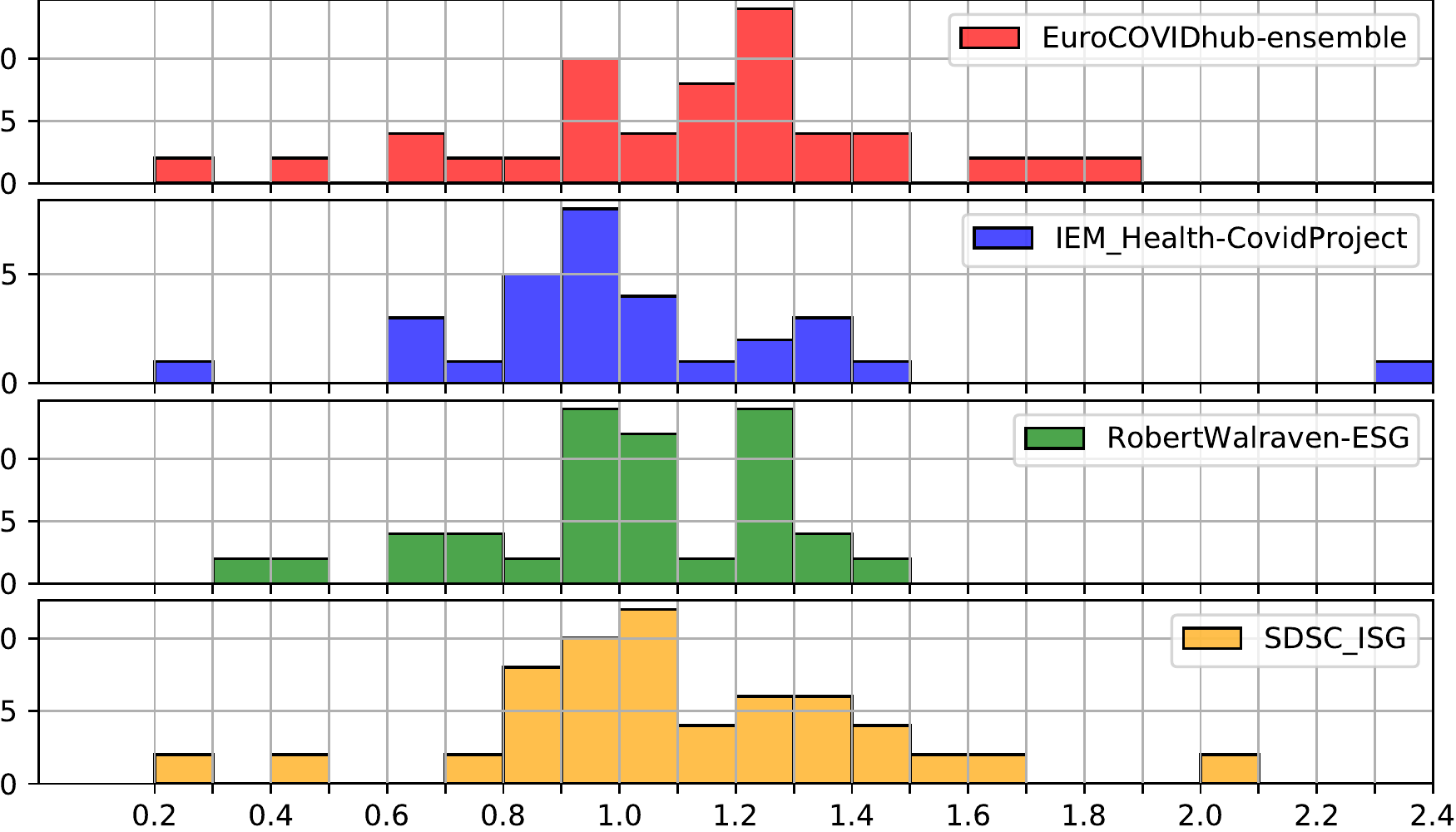}
    \caption{Histograms for the MAE (in x-axis) based on  1 week ahead deaths forecasts for 31 European countries}
    \label{fig:deaths_w1_MAE}
\end{figure}

\begin{figure}
    \centering
    \includegraphics[width=0.7\textwidth]{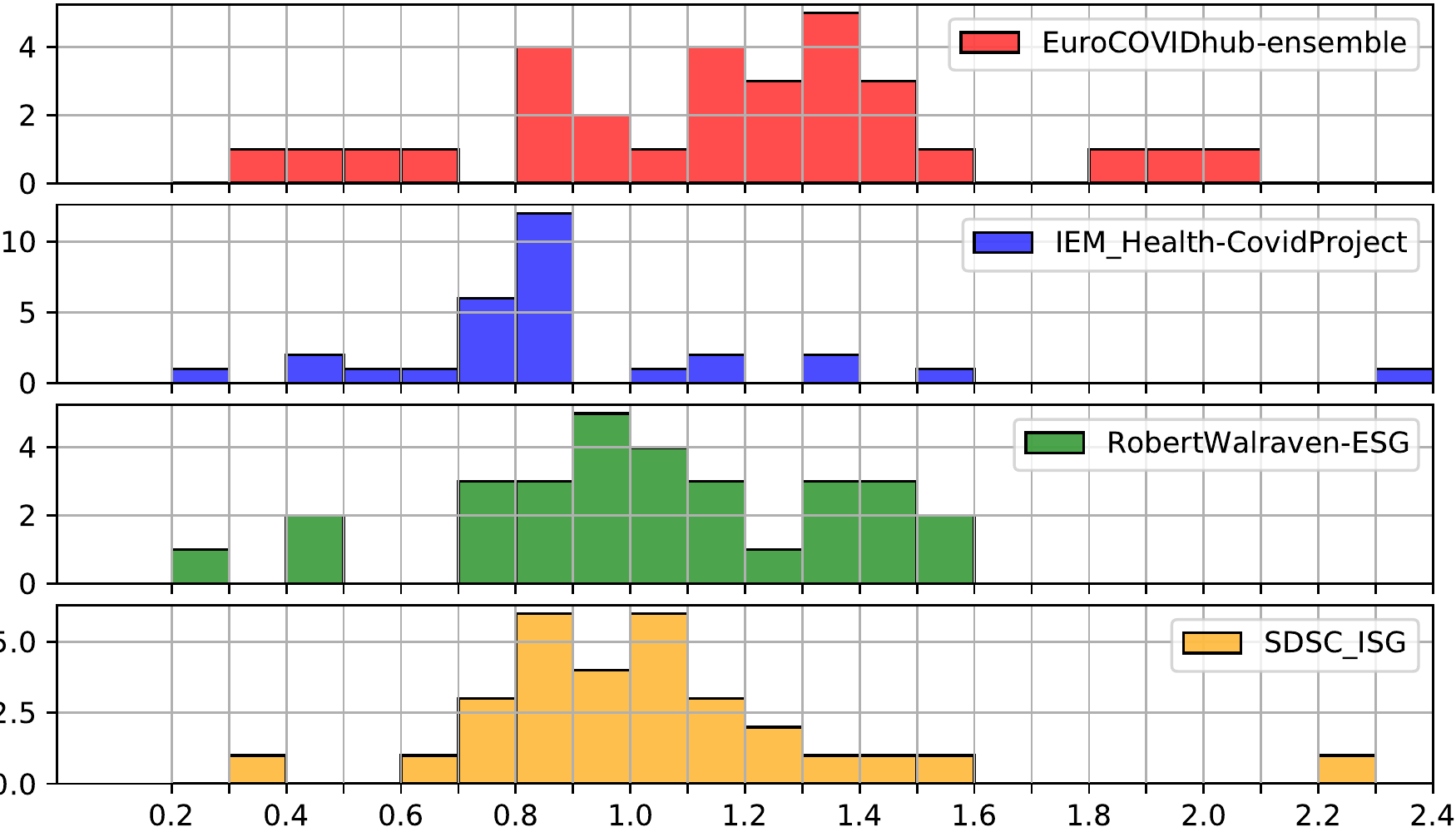}
    \caption{Histograms for the average WIS (in x-axis) based on 1 week ahead deaths  forecasts  for 31 European countries}
    \label{fig:deaths_w1_WIS}
\end{figure}

\begin{figure}
    \centering
    \includegraphics[width=0.7\textwidth]{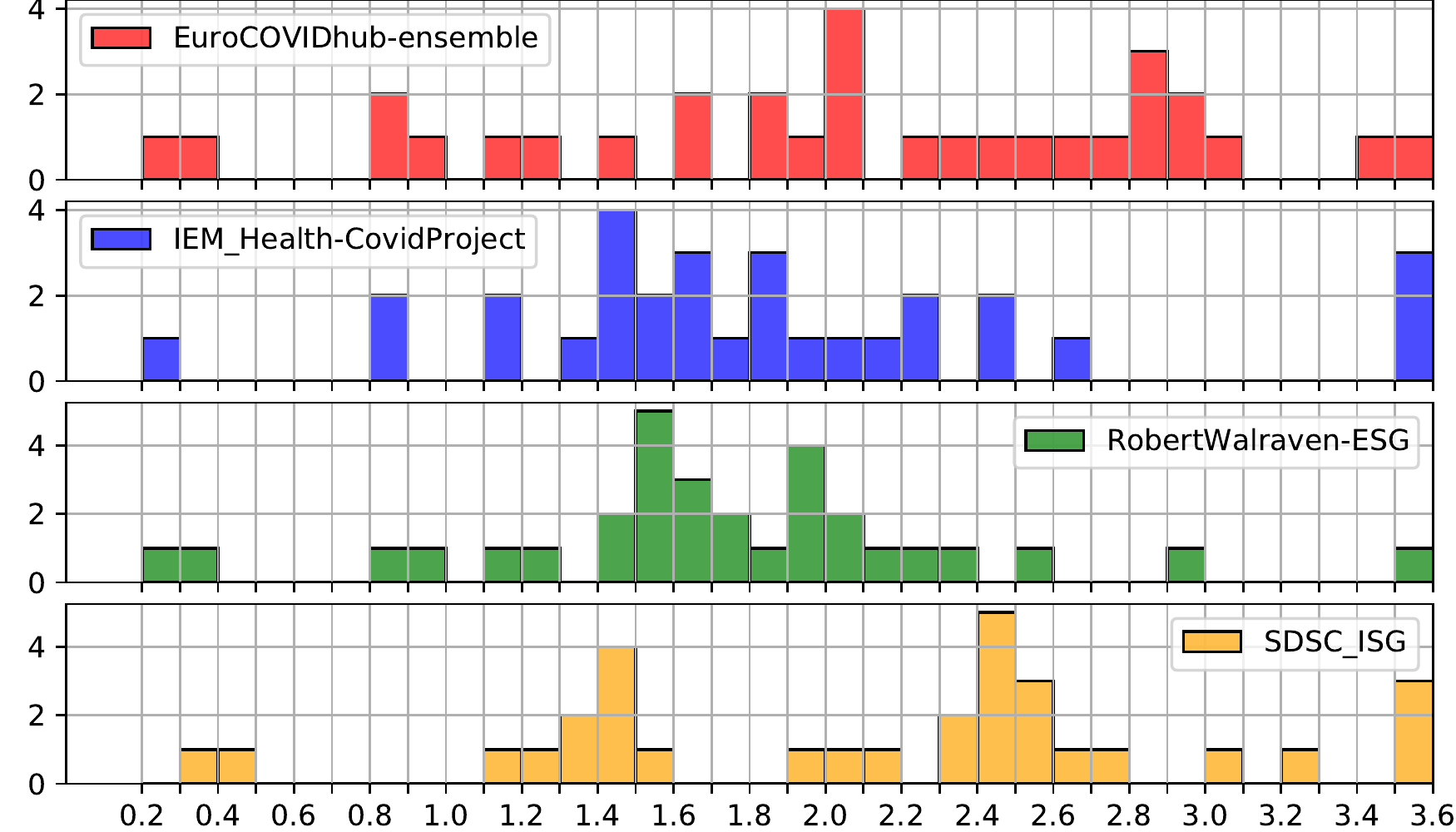}
    \caption{Histograms for the MAE (in x-axis) based on 2 week ahead deaths forecasts for 31 European countries}
    \label{fig:deaths_w2_MAE}
\end{figure}

\begin{figure}
    \centering
    \includegraphics[width=0.7\textwidth]{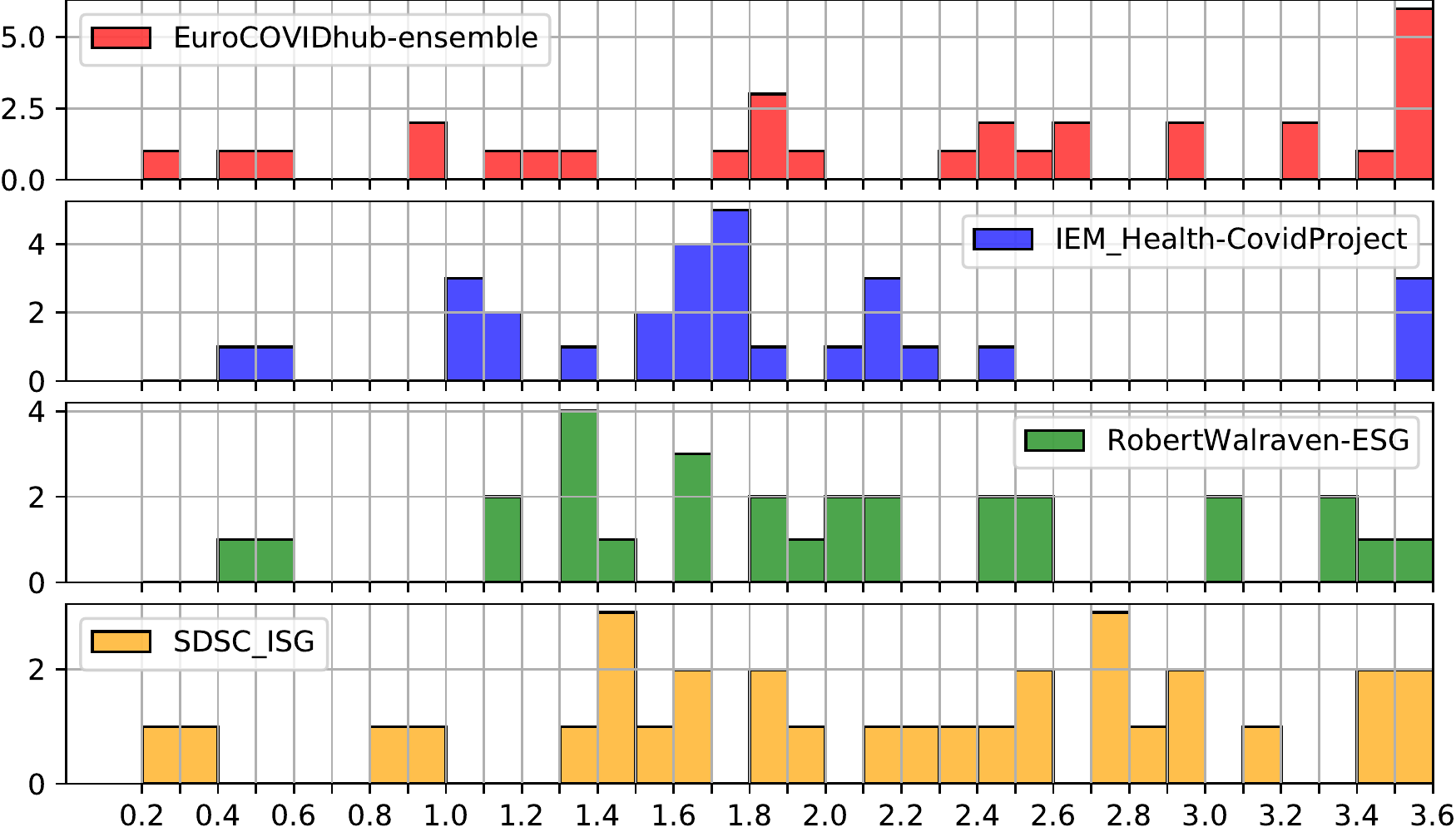}
    \caption{Histograms for the average WIS (in x-axis) based on 2 week ahead deaths  forecasts  for 31 European countries}
    \label{fig:deaths_w2_WIS}
\end{figure}

\textcolor{highlight1}{We provide a brief comparison of the performance of our strategy to forecast deaths (individually in the same way as cases) with a few methods from the European forecast Hub  for 31 European countries in a similar way to what we did for cases. 
We identified two forecasting methods IEM\_Health-CovidProject and RobertWalraven-ES, which were regularly submitting to the European COVID forecast Hub apart from  EuroCOVIDhub-ensemble and EuroCOVIDhub-baseline. The results demonstrate that for 1-week ahead forecast our methodology obtains levels of performance comparable to those of other methods submitted to EU COVID Forecast Hub: on Fig.~\ref{fig:deaths_w1_MAE}, \ref{fig:deaths_w2_MAE} and Fig.~\ref{fig:deaths_w1_WIS}, \ref{fig:deaths_w2_WIS} one can see that mean absolute error (MAE) and WIS  normalized by the respective errors of the EU COVID hub baseline of our method (SDSC\_ICG) are aligned with the other methods.  
}
\section*{B. Raw data preprocessing }

In this section, we describe the preliminary data cleaning steps and smoothing for the daily cases and deaths. 

\subsection*{Negative values}
 
When a negative count is reported following a reassessment of previously reported cases or deaths, we substitute the negative value with the estimate $x_t$ computed from the daily observations a week before multiplied by the growth factor computed from two weekly observations: during a week before and two weeks before the negative value had occurred, i.e. with
\textcolor{highlight}{
$$
x_t \frac{X_{t-1}}{X_{t-8}}
$$
}
where $x_t$  and $X_t=\sum_{k=0}^6 x_{t-k}$ are daily and weekly observations respectively.
Next we reduce the counts in the whole previous history by a constant multiplicative factor $c$ to match  the cumulative counts obtained by removing this negative quantity from the cumulative counts. That is we compute $c=(\sum_{s=0}^t x_t)/ (\sum_{s=0}^{t-1} x_t)$ and $x_t$ is updated to $x_t \leftarrow c\, x_t.$
For the case of large and  significantly delayed reports or reassessment leading symmetrically to a large positive spike in the data, this is ignored at this stage, and will be addressed by the trend-estimation method.

\subsection*{Identifying last missing daily reports}
One of the difficulties with the data sources that we are using is that no distinction is made between a missing report and an existing report stating that no new cases should be reported on a given day: in both cases the database contains a zero. This is of course because, in practice, there is often no distinction made in the reporting protocols. It is however important for our models to be able to distinguish between these two situations. To distinguish missing values from actual zeros, we proceed as follows:  if the last observation in the data is zero, we compute an estimate of the Poisson rate by taking the average of observations during the week before zero occurs. If the probability of observing zero new cases given this estimated rate is too low, we consider the zero value to be a missing report one and exclude it from the history for further trend estimation and forecasting. The forecasting starts from the next day from the last initial observation. 

\subsection*{Imputations}
Many countries have seasonal patterns in which no data is reported on certain days, typically during the weekend, and where all the new cases that appeared during these days are reported all together on the next reporting day (typically a Monday). We use as a preprocessing a simpler imputation scheme, which consists in reassigning the data declared on that last day uniformly over the previous days of missing report and the following reporting day.

\section*{C. Details of the piecewise STL algorithm}

Since the seasonal pattern might evolve over time due to changes in the reporting pattern we propose to apply STL in a piecewise fashion. This allows our method to better adapt to changes in the seasonal component, as the hyperparameters defining the smoothing levels can then change in each separate segment modelled. \textcolor{highlight}{
To estimate the trend locally in the whole period of observations, we split the observed time interval into half-overlapping intervals of 6 weeks. These time intervals are defined from the end of the time series backwards, so that the time series ends with the last segment of 6 months.\\
First, we apply STL to estimate the trend of the last subinterval $[T-L+1,T]$, where $L=42$ (corresponding to 4 weeks). For the last subinterval, the STL trend is rescaled to preserve the number of observations in the last $L/2$ days to obtain the estimate $s_{-1}$ in $[T-L+1,T]$.
Next, the trend estimation proceeds as follows. For the two overlapping subintervals, e.g.\ consider $[T-L+1,T]$ and $[T-3L/2+1,T-L/2]$, we take the estimate $s_{-1}$ and we estimate the trend $\tilde{s}$ in  $[T-3L/2+1, T-L/2]$. In order to smoothly join two local trends $s_{-1}$ and $\tilde{s}_{-2}$ in the interval $[T-L+1,T-L/2]$ we use a simple weighted interpolation and obtain the trend in $[T-3L/2+1,T]$ :
\begin{equation*}
s_{-2}(t_0+\tau) = 
    \begin{cases}
        \tilde{s}(t_0+\tau), & \quad \tau=-L/2+1,\dots,0,\\  
        \sigma(\tau)\,
        \tilde{s}(t_0+\tau) + (1-\sigma(\tau))\, s_{-1}(t_0+\tau) , & \quad \tau=1,\dots,L/2,\\   s_{-1}(t_0+\tau), & \quad \tau=L/2+1,\dots,L,\\ 
    \end{cases}
\end{equation*} 
where $t_0=T-L+1$, and $\sigma(\tau) = (1+\exp(a(\tau-1)-b))^{-1}$ with $a=21.1/L$, $b=5.46$. 
We additionally apply rescaling to redistribute the possible outliers, removed by trend estimation: we compare the sum of the numbers so far estimated by the trend, e.g. $S_{-2} = \sum_{i=0}^{3L/2-1}\tilde{s}_{-2}(T-i)$, with the corresponding number of raw daily observations $\kappa_{-2}= \sum_{i=0}^{3L/2-1} x_{T-i}$: if the excess $\kappa_{-2}-S_{-2}$ is positive, it is added to the observations before $T-3L/2+1$ by rescaling, otherwise the trend is rescaled such that the sum of estimated numbers meets $\kappa_{-2}$. The procedure continues with the next local trend estimate ($\tilde{s}$) from the corrected data. Note that $\kappa$ is always computed from raw observations. Local trend estimation with rescaling repeats backward until we reach the beginning of the time interval. As a result, we get a smooth trend, the sum of which is equal to the sum of raw daily observations.
}

\section*{D. Outlier detection scheme}

One of the criteria for the inclusion of a country into our main evaluation set is that there are not too many large delayed reports. We assimilate these as outliers and use a simple estimate for each time series of the number of outliers. The corresponding outlier detection scheme is based on the Median Absolute Deviation (MAD), which is defined as ${\rm MAD} = {\rm median} (|x_i - {\rm median}(x_i)|)$  for daily observations $x_i$.
For each country, we detect outliers using a sliding window MAD estimate.  More precisely, for each daily value, we compute the MAD in a symmetric window of 22 days around that day. For the MAD to be a consistent estimator for the standard deviation, we multiply it by a constant scale factor of 1.4826, which relates the MAD to the standard deviation for a Gaussian distribution. If the daily value differs from the median in the window by more than 2 standard deviations, we consider it as an outlier. Note that this procedure is only used for the construction of the set of countries in the evaluation set, and not for the trend estimation or forecast.

\section*{E. Total Coverage}
 
\textcolor{highlight}{
We define the \emph{total coverage} of a probabilistic forecast $P$ the sum over all levels considered of the coverage of the intervals $[q_k,q_{2K+2-k}]$ as defined in~\cite{cramer2021evaluation}: 
$$C(P,A,\xi) = \sum_{k=1}^K \mathbbm{1}\{q_k\leq \xi \leq q_{2K+2-k}\},$$
where $\mathbbm{1}\{\cdot\}$ is $1$ if the condition is satisfied and zero otherwise.
We the define the \emph{mean total coverage} as 
$${\rm MC}=\frac{1}{T} \sum_{t=1}^T C(P_t,A,X_t),$$
where $X_t$ are the weekly total number of cases/deaths.
}
\section*{F. Growth rate analysis}
\begin{figure}
\centering
\includegraphics[width=\textwidth]{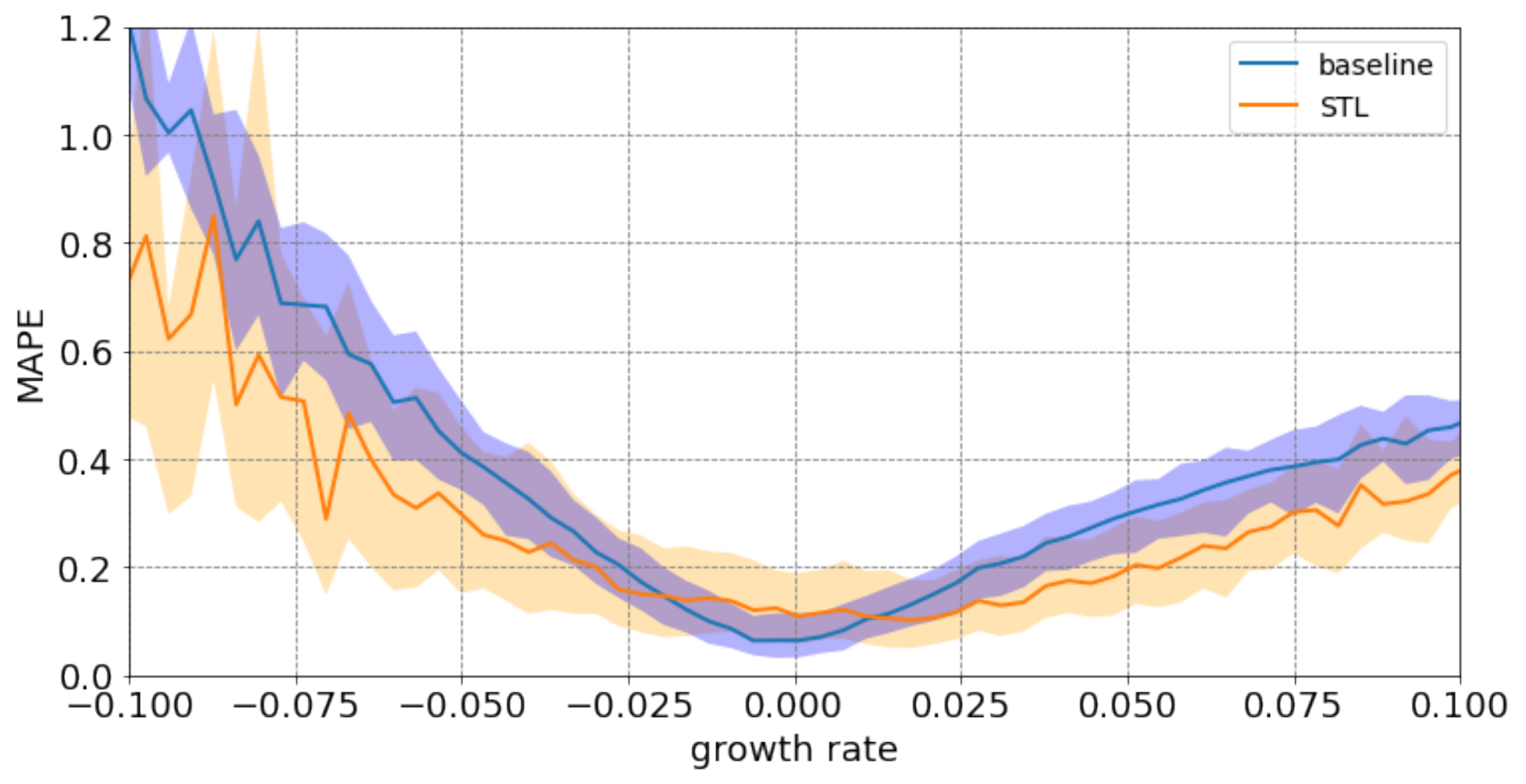}
\caption{Dependence of the error on the relative slope of the trend: Median (solid line) and interquartiles (shaded region) over all countries (this excludes 11 countries with more than 90\% of zero daily observations: Marshall Islands, Grenada, Vanuatu, Tanzania, Fiji, Saint Kitts and Nevis, Micronesia, Samoa, the Holy See, Solomon Islands, Laos) of the MAPE of each forecasting algorithm (blue: baseline, orange: proposed forecast) as a function of the growth rate (aka relative slope) of the trend.  Since the baseline assumes zero slope it has lower median error when the absolute growth rate is less than 3\%, but larger median error otherwise.}
\label{fig:slope}
\end{figure}

 \textcolor{highlight}{
 To be able to estimate the growth rate of the trend, we first compute an independent estimate of the trend using cubic B-splines on the weekly data and next compute the growth rate as the slope of the trend normalized by the trend value. To aggregate AE values for different countries, we use the MAPE (the mean of ${\rm AE}(F_t)/X_t$) on the weekly forecasts  in the evaluation period instead of the MAE, to bring the errors for each country on a comparable scale. Given that the growth rate as a measure of the slope is comparable across countries, we pool the data from all 
countries to obtain 
Fig.~\ref{fig:slope}. The baseline 
performs best when there are no changes in the number of cases/deaths (i.e., the growth rate of 0 or constant trend). However, our method outperforms the baseline predictor as soon as the growth rate is larger than 3\% in absolute value, which shows that the proposed forecast is informative as soon as the trend is not flat.
}

\newpage


\newpage

\newpage

\newpage

\newpage

\newpage

\newpage

\newpage

\newpage

\begin{figure}
    \centering
    \includegraphics[width=0.99\textwidth]{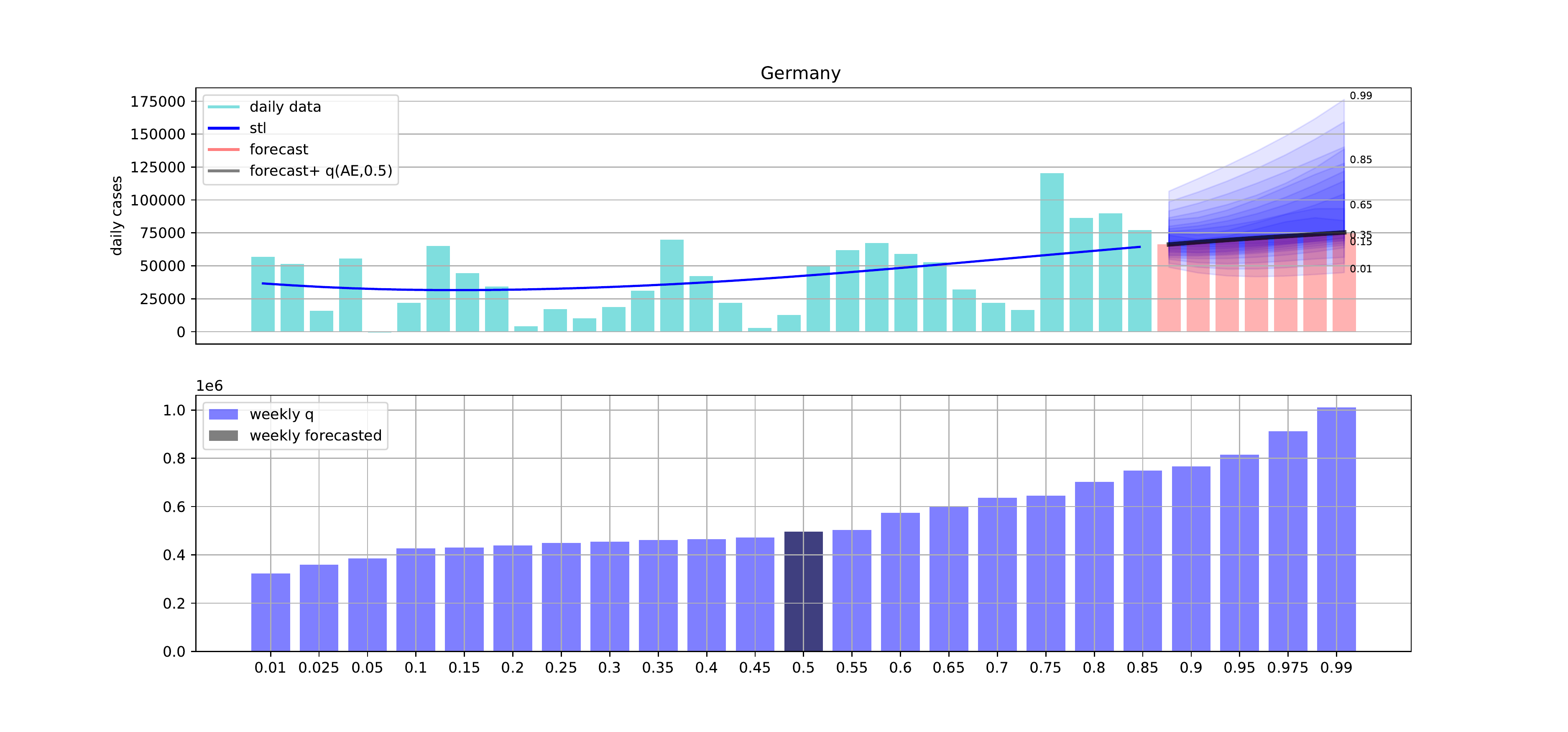}
    \caption{Confidence intervals for Germany for one week ahead prediction obtained on January 14, 2022. The upper subplot demonstrates the forecast together with the predictive intervals; additional spline smoothing is applied for each confidence level to smooth the quantiles in time. Lower plot shows the estimated quantiles for the weekly number.}
    \label{ci:us}
\end{figure}

\begin{figure}
    \centering
    \includegraphics[width=0.99\textwidth]{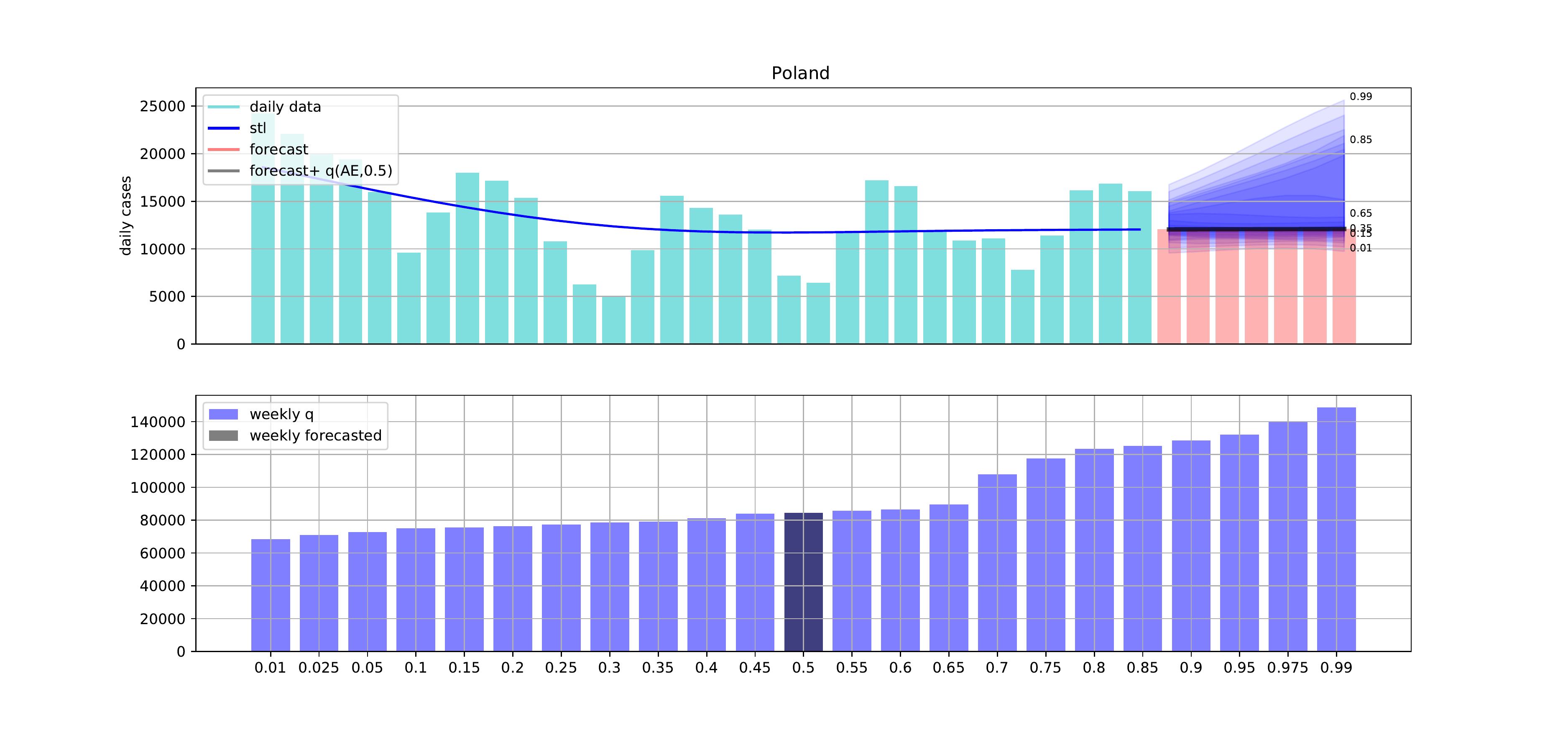}
    \caption{Confidence intervals for Poland for one week ahead prediction obtained on January 14, 2022. The upper subplot demonstrates the forecast together with the predictive intervals; additional spline smoothing is applied for each confidence level to smooth the quantiles in time. Lower plot shows the estimated quantiles for the weekly number.}
    \label{ci:poland}
\end{figure}

\begin{figure}
    \centering
    \includegraphics[width=0.99\textwidth]{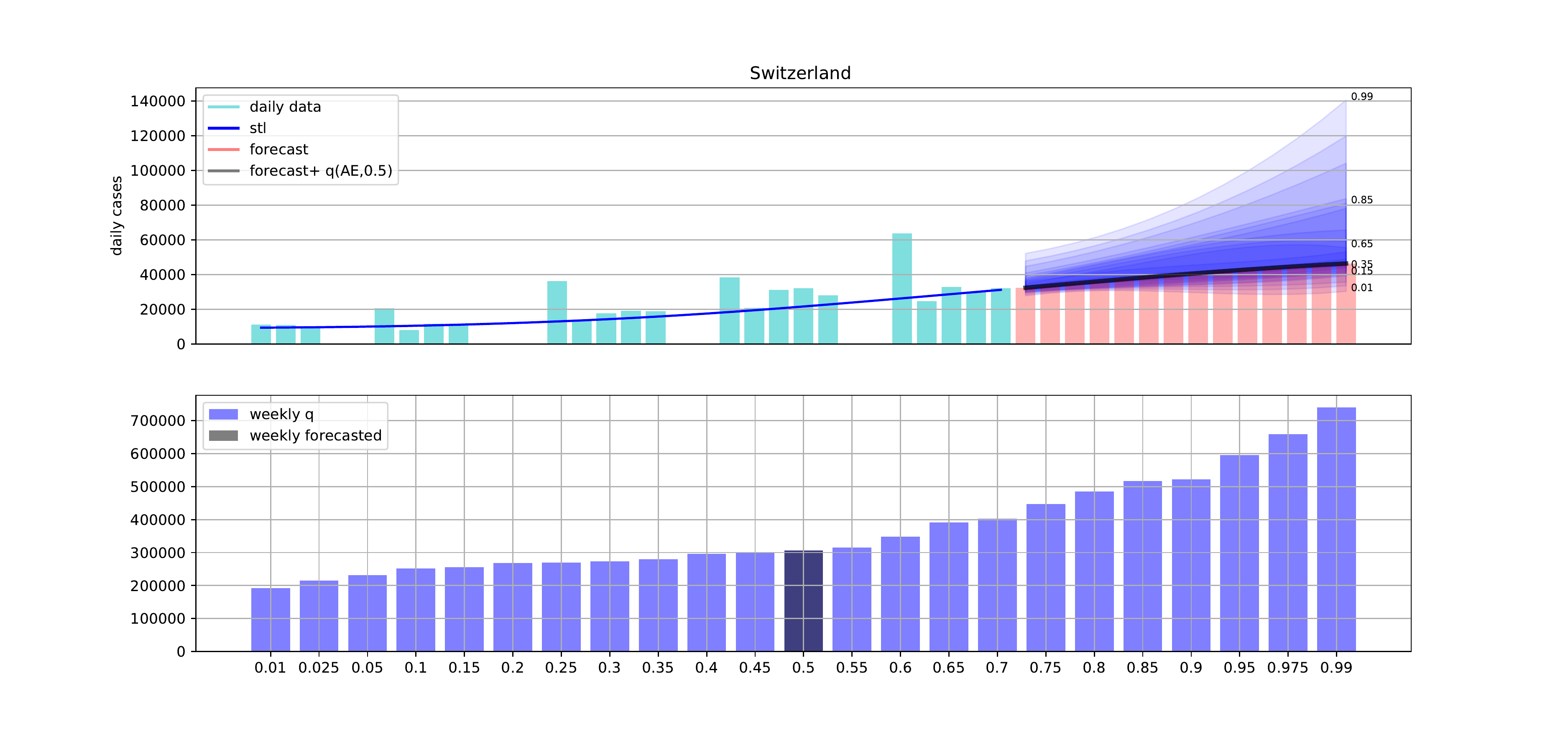}
    \caption{Confidence intervals for Switzerland for two weeks ahead prediction obtained on January 14, 2022. The upper subplot demonstrates the forecast together with the predictive intervals; additional spline smoothing is applied for each confidence level to smooth the quantiles in time. Lower plot shows the estimated quantiles for weekly numbers of the second forecasted week.}
    \label{ci:sw}
\end{figure}

\newpage


\begin{table}[]
 \centering
\begin{tabular}{lrrrrrrr}
\toprule
       country &                     EUHub-ens&                     IEM\_Health &           ILM &               MUNI&                        RW &                  USC &       SDSC\_ISG \\
\midrule
       Austria &                         0.71 &                           1.05 &           0.79 &              0.82 &                  {1.05}& \textbf{\textcolor{orange}{0.64}}&            0.70 \\
       Belgium &                         0.81 &                           1.11 &           0.85 &              1.09 &                  {1.33}& \textbf{\textcolor{orange}{0.77}}&            1.49 \\
      Bulgaria &                         0.77 &                     1.34 &           1.14 & \textbf{\textcolor{orange}{0.53}}&                      1.13 &                 0.80 &            0.79 \\
       Croatia &                         0.70 &                           1.16 &           0.94 &              0.72 &                      0.94 &                 0.83 & \textbf{\textcolor{orange}{0.61}} \\
        Cyprus &                         0.85 &        \textbf{\textcolor{orange}{0.77}}&           2.28 &              0.93 &                      1.13 &                 0.99 &            0.81 \\
       Czechia &                         0.65 &                           0.89 &           0.74 &              0.75 &                   {1.02}&\textbf{\textcolor{orange}{0.61}}&            0.62 \\
       Denmark &                         0.74 &                           0.76 &           0.98 &              0.94 &                      1.00 &                 0.89 & \textbf{\textcolor{orange}{0.64}} \\
       Estonia &                         0.88 &                           0.85 &           1.16 &              1.02 &                      1.06 &                 1.24 & \textbf{\textcolor{orange}{0.81}} \\
       Finland &                         0.84 &                           0.82 &           1.02 &              1.10 &                      1.18 &                 1.17 & \textbf{\textcolor{orange}{0.80}} \\
        France &                         0.61 &                           0.70 &           0.73 &              0.78 &                      0.93 &                 0.66 & \textbf{\textcolor{orange}{0.54}} \\
       Germany &      \textbf{\textcolor{orange}{0.60}}&                           1.19 &           0.65 &              0.78 &                      0.89 &                 0.74 &            0.69 \\
        Greece &                         0.96 &                       1.44 &         1.46 & \textbf{\textcolor{orange}{0.84}}&                      1.34 &                 1.47 &            0.94 \\
       Hungary &                         0.66 &                           1.06 &     0.79 & \textbf{\textcolor{orange}{0.58}}&                      1.10 &                 0.65 &            0.64 \\
       Iceland &                         0.65 &        \textbf{\textcolor{orange}{0.43}}&           2.18 &              0.90 &                      0.57 &                 0.72 &            0.79 \\
       Ireland &                         1.25 &                           1.45 &           1.98 &              1.27 &                  {1.40}& \textbf{\textcolor{orange}{1.20}}&            1.31 \\
         Italy &                         0.52 &                     0.63 &           0.60 & \textbf{\textcolor{orange}{0.38}}&                      1.08 &                 0.69 &            0.47 \\
        Latvia &                         0.86 &                           0.99 &           1.04 &              1.00 &                  {1.07}& \textbf{\textcolor{orange}{0.68}}&            0.73 \\
 Liechtenstein &      \textbf{\textcolor{orange}{0.71}}&                           1.12 &           1.52 &              0.86 &                      0.83 &                 1.24 &            0.95 \\
     Lithuania &                         0.64 &                     0.84 &           1.13 & \textbf{\textcolor{orange}{0.48}}&                      0.92 &                 0.98 &            0.64 \\
    Luxembourg &      \textbf{\textcolor{orange}{0.89}}&                           1.14 &           1.24 &              1.06 &                      1.16 &                 1.37 &            1.07 \\
         Malta &                         1.10 &                     1.52 &           5.11 & \textbf{\textcolor{orange}{0.74}}&                      1.21 &                 1.45 &            1.20 \\
   Netherlands &                         0.69 &                      0.81 &           1.09 &\textbf{\textcolor{orange}{0.56}}&                      1.07 &                 1.00 &            0.84 \\
        Norway &                         0.70 &                           0.71 &           0.86 &              0.62 &                      1.06 &                 0.94 & \textbf{\textcolor{orange}{0.54}} \\
        Poland &                         0.49 &                      0.92 &           0.64 &\textbf{\textcolor{orange}{0.29}}&                      1.00 &                 0.73 &            0.53 \\
      Portugal &                         0.70 &                           0.79 &           0.97 &              0.93 &                      1.04 &                 1.01 & \textbf{\textcolor{orange}{0.63}} \\
       Romania &                         0.55 &                      0.97 &           0.57 &\textbf{\textcolor{orange}{0.33}}&                      0.93 &                 0.44 &            0.52 \\
      Slovakia &                         0.70 &                  0.81 & \textbf{\textcolor{orange}{0.67}}&              0.67 &                      0.91 &                 0.79 &            0.77 \\
      Slovenia &                         0.79 &                     1.25 &           1.18 & \textbf{\textcolor{orange}{0.48}}&                      1.13 &                 1.06 &            0.59 \\
         Spain &                         0.79 &                           0.96 &           0.98 &              0.98 &                      0.90 &                 0.99 & \textbf{\textcolor{orange}{0.60}} \\
        Sweden &                         0.87 &                           1.00 &           0.90 &              1.24 &                      0.86 &                 0.69 & \textbf{\textcolor{orange}{ 0.66}} \\
   Switzerland &     \textbf{\textcolor{orange}{0.67}}&                           0.67 &           0.81 &              0.89 &                      0.95 &                 0.72 &            0.70 \\
\bottomrule
    ranks best in   &                               4 &                                2  &           1 &                 10 &                         0 &                    5 &               9 \\ 
    in top 2  &             15 &                                5  &           2 &                 13 &                         2 &                    7 &              18 \\ 
    in top 3  &              28 &                                8  &           3 &                 16 &                         4 &                    9 &              25 \\ 
    in top 4  &              31 &                                14  &          8 &                 20 &                         6 &                   16 &              29 \\ 

\bottomrule
\end{tabular}
    \caption{One week ahead forecast AE normalized by the EuroCovidhub\_baseline AE. The values, which are highlighted in bold and orange color, correspond to the best performance. The lower part of the table reports for each method the number of countries for which its forecast is best, or in the top $2$, $3$ or $4$ best performing methods.}
    \label{tab:AE_w1}
\end{table}

\begin{table}[]
    \centering
\begin{tabular}{lrrrrrrr}
\toprule
       country &  EUHub-ens  &  IEM\_Health  & ILM& MUNI   &  RW  &  USC   &  SDSC\_ISG \\
\midrule
       Austria &                      0.55 &                        0.99 &        0.54 &           0.78 &                   1.06 &              0.61 & \textbf{\textcolor{orange}{0.54}} \\
       Belgium &    \textbf{\textcolor{orange}{0.61}} &                        0.92 &        0.64 &           0.97 &                   1.30 &              0.79 &             1.21 \\
      Bulgaria &                      0.57 &                        1.11 &        0.88 & \textbf{\textcolor{orange}{0.45}} &                   1.06 &              0.80 &             0.66 \\
       Croatia &                      0.58 &                        0.83 &        0.73 & \textbf{\textcolor{orange}{0.56}} &                   0.81 &              0.85 &             0.58 \\
        Cyprus &                      0.66 &                        0.77 &        2.19 &           0.83 &                   1.11 &              1.04 &\textbf{\textcolor{orange}{0.59}} \\
       Czechia &                      0.49 &                        0.67 &        0.58 &           0.68 &                   1.11 &              0.46 &\textbf{\textcolor{orange}{0.45}} \\
       Denmark &                      0.59 &                        0.69 &        0.77 &           0.84 &                   1.08 &              0.99 &\textbf{\textcolor{orange}{0.50}} \\
       Estonia &                      0.77 &      \textbf{\textcolor{orange}{ 0.73}} &        0.88 &           1.03 &                   1.07 &              1.03 &             0.77 \\
       Finland &                      0.64 &        \textbf{\textcolor{orange}{0.59}} &        0.73 &           1.01 &                   1.22 &              1.32 &             0.64 \\
        France &   \textbf{\textcolor{orange}{0.45}} &                        0.60 &        0.65 &           0.65 &                   1.00 &              0.76 &             0.47 \\
       Germany &                      0.46 &                        0.86 & \textbf{\textcolor{orange}{0.38}} &           0.57 &                   0.77 &              0.68 &             0.47 \\
        Greece &                      0.76 &                        1.18 &        1.15 &\textbf{\textcolor{orange}{0.69}} &                   1.39 &              1.38 &             0.86 \\
       Hungary &                      0.50 &                        0.90 &        0.54 & \textbf{\textcolor{orange}{0.47}} &                   1.14 &              0.61 &             0.54 \\
       Iceland &                      0.73 &                        1.30 &        1.42 &           0.57 &   \textbf{\textcolor{orange}{0.38}} &              1.41 &             1.22 \\
       Ireland &    \textbf{\textcolor{orange}{1.03}} &                        1.17 &        1.64 &           1.10 &                   1.38 &              1.37 &             1.25 \\
         Italy &                      0.41 &                        0.64 &        0.51 & \textbf{\textcolor{orange}{0.31}} &                   1.09 &              0.98 &             0.39 \\
        Latvia &                      0.62 &                        0.73 &        0.69 &           0.96 &                   0.94 &              0.58 & \textbf{\textcolor{orange}{0.54}} \\
 Liechtenstein &                      0.68 &                        0.81 &        1.09 & \textbf{\textcolor{orange}{0.65}} &                   0.83 &              1.24 &             0.73 \\
     Lithuania &                      0.49 &                        0.72 &        0.95 & \textbf{\textcolor{orange}{0.43}} &                   0.98 &              0.98 &             0.61 \\
    Luxembourg &       \textbf{\textcolor{orange}{0.73}} &                        1.27 &        1.21 &           0.96 &                   1.39 &              1.58 &             1.12 \\
         Malta &                      0.82 &                        1.75 &        5.07 & \textbf{\textcolor{orange}{0.58}} &                   1.37 &              1.75 &             1.31 \\
   Netherlands &      \textbf{\textcolor{orange}{0.55}} &                        0.85 &        1.11 &           0.55 &                   1.15 &              0.82 &             0.83 \\
        Norway &                      0.55 &                        0.57 &        0.64 &           0.47 &                   0.99 &              0.84 &\textbf{\textcolor{orange}{0.35}} \\
        Poland &     0.35 &                        0.69 &        0.40 &\textbf{\textcolor{orange}{0.23}}&                   0.91 &              0.57 &             0.42 \\
      Portugal &       \textbf{\textcolor{orange}{0.57}} &                        0.64 &        0.73 &           1.05 &                   1.08 &              1.23 &             0.62 \\
       Romania &                      0.36 &                        0.81 &        0.36 &   \textbf{\textcolor{orange}{0.25}} &                   0.86 &              0.38 &             0.34 \\
      Slovakia &                      0.54 &                        0.57 &\textbf{\textcolor{orange}{0.47}} &           0.60 &                   0.87 &              0.75 &             0.61 \\
      Slovenia &                      0.64 &                        0.95 &        0.91 & \textbf{\textcolor{orange}{0.37}} &                   1.13 &              0.88 &             0.54 \\
         Spain &                      0.70 &                        0.97 &        1.27 &           1.15 &                   1.13 &              1.22 &\textbf{\textcolor{orange}{0.52}} \\
        Sweden &                      0.78 &                        1.01 &        0.80 &           1.39 &                   0.90 &              0.90 & \textbf{\textcolor{orange}{0.62}} \\
   Switzerland &      \textbf{\textcolor{orange}{0.53}} &                        0.66 &        0.65 &           0.81 &                   1.09 &              0.78 &             0.55 \\
\bottomrule
    ranks best in   &           7 &                       2  &           2 &                 11 &                         1 &                    0 &               8 \\ 
    in top 2  &             21 &                                2  &           4 &                 16 &                         1 &                    2 &              16 \\ 
    in top 3  &              31 &                                9  &           7 &                 16 &                         1 &                    4 &              25 \\ 
    in top 4  &              31 &                                14  &          19 &                 20 &                        3 &                    8 &              29 \\ 
 \bottomrule
\end{tabular}
    \caption{One week ahead forecast WIS normalized by the EuroCovidhub\_baseline WIS. The values, which are highlighted in bold and orange color, correspond to the best performance. The lower part of the table reports for each method the number of countries for which its forecast is best, or in the top $2$, $3$ or $4$ best performing methods.}
    \label{tab:WIS_w1}
\end{table}

\begin{table}[]
    \centering 
\begin{tabular}{lrrrrrrr}
\toprule
       country &  EUHub-ens  &  IEM\_Health  & ILM& MUNI   &  RW  &  USC   &  SDSC\_ISG \\
\midrule
       Austria &                         0.81 &                           1.49 &           1.43 &              0.60 &                      1.00 & \textbf{\textcolor{orange}{0.56}} &            0.93 \\
       Belgium &   \textbf{\textcolor{orange}{0.74}} &                           1.45 &           1.31 &              1.21 &                      1.21 &                 0.91 &            1.23 \\
      Bulgaria &                         0.95 &                           1.59 &           1.68 &\textbf{\textcolor{orange}{0.62}} &                      1.05 &                 0.87 &            1.04 \\
       Croatia &                         0.88 &                           1.66 &           1.56 &\textbf{\textcolor{orange}{0.74}} &                      0.98 &                 0.94 &            0.92 \\
        Cyprus &                         1.04 &                           0.88 &           3.69 & \textbf{\textcolor{orange}{0.78}} &                      1.13 &                 0.91 &            0.94 \\
       Czechia &                         0.99 &                           1.00 &           1.65 &              0.71 &                      0.89 &                 1.24 &            \textbf{\textcolor{orange}{0.68}} \\
       Denmark &                         0.69 &                           0.86 &           1.30 &              1.04 &                      1.00 &                 0.97 &            \textbf{\textcolor{orange}{0.65}} \\
       Estonia &                         0.82 &             \textbf{\textcolor{orange}{0.82}} &           1.32 &              0.92 &                      0.97 &                 1.33 &            0.88 \\
       Finland &    \textbf{\textcolor{orange}{0.88}} &                           0.91 &           1.23 &              1.11 &                      1.35 &                 1.20 &            0.90 \\
        France &\textbf{\textcolor{orange}{0.64}} &                           0.86 &           1.36 &              0.78 &                      0.96 &                 0.78 &            0.64 \\
       Germany &                         0.78 &                           1.49 &           1.16 &              0.79 &                      0.87 &  \textbf{\textcolor{orange}{0.56}} &            0.76 \\
        Greece &                         1.34 &                           1.71 &           1.96 &\textbf{\textcolor{orange}{0.90}} &                      1.43 &                 1.44 &            1.15 \\
       Hungary &                         0.99 &                           1.57 &           1.52 &              0.74 &                      1.03 & \textbf{\textcolor{orange}{0.69}} &            0.71 \\
       Iceland &                         0.47 &          \textbf{\textcolor{orange}{0.41}} &           2.73 &              1.00 &                      0.95 &                 0.92 &            1.00 \\
       Ireland &                         1.51 &                           1.78 &           3.32 &\textbf{\textcolor{orange}{1.35}} &                      1.48 &                 1.43 &            1.58 \\
         Italy &    \textbf{\textcolor{orange}{0.54}} &                           0.77 &           0.99 &              0.56 &                      1.07 &                 0.77 &            0.56 \\
        Latvia &                         0.92 &                           1.25 &           1.32 &              0.99 &                      0.93 & \textbf{\textcolor{orange}{0.62}} &            0.79 \\
 Liechtenstein &      \textbf{\textcolor{orange}{0.85}} &                           1.75 &           1.38 &              0.96 &                      0.91 &                 1.77 &            1.01 \\
     Lithuania &                         0.61 &                           1.03 &           1.51 &\textbf{\textcolor{orange}{0.52}} &                      0.91 &                 0.92 &            0.78 \\
    Luxembourg &        \textbf{\textcolor{orange}{0.85}} &                           1.07 &           1.03 &              1.05 &                      1.10 &                 2.09 &            1.04 \\
         Malta &                         1.67 &                           1.90 &           9.98 &\textbf{\textcolor{orange}{0.77}} &                      1.13 &                 1.69 &            1.40 \\
   Netherlands &                         1.28 &                           0.89 &           3.01 &\textbf{\textcolor{orange}{0.86}} &                      1.04 &                 1.05 &            0.89 \\
        Norway &                         0.85 &                           0.93 &           1.12 &              0.69 &                      1.09 &                 1.43 &            \textbf{\textcolor{orange}{0.66}} \\
        Poland &                         0.79 &                           1.29 &           1.37 &\textbf{\textcolor{orange}{0.43}} &                      1.01 &                 1.03 &            0.49 \\
      Portugal &                         0.82 &                           0.91 &           1.31 &              0.97 &                      1.08 &                 1.07 &            \textbf{\textcolor{orange}{0.74}} \\
       Romania &                         0.73 &                           1.28 &           1.11 &              0.52 &                      0.86 &\textbf{\textcolor{orange}{0.46}} &            0.71 \\
      Slovakia &        \textbf{\textcolor{orange}{0.54}} &                           0.63 &           0.56 &              0.62 &                      0.88 &                 0.66 &            0.71 \\
      Slovenia &                         1.15 &                           1.56 &           1.70 &\textbf{\textcolor{orange}{0.58}} &                      1.08 &                 1.66 &            0.92 \\
         Spain &                         1.00 &                           1.11 &           1.70 &              0.96 &                      0.94 &                 1.10 &           \textbf{\textcolor{orange}{0.82}} \\
        Sweden &                         0.77 &                           1.04 &           0.85 &              1.19 &                      0.84 & \textbf{\textcolor{orange}{0.60}} &            0.61 \\
   Switzerland &       \textbf{\textcolor{orange}{0.67}} &                           1.02 &           1.19 &              0.91 &                      1.01 &                 0.82 &            0.71 \\
\bottomrule

    ranks best in   &                               8 &                                2  &           0 &                 10 &                         0 &                    6 &               5 \\ 
    in top 2  &             14 &                                3  &           2 &                 14 &                         3 &                    9 &              17 \\ 
    in top 3  &              22 &                                7  &           2 &                 19 &                         7 &                    13 &              23 \\ 
    in top 4  &              29 &                                10  &          2 &                 27 &                        14 &                    15 &              27 \\ 
 \bottomrule   
\end{tabular}
    \caption{Two week ahead forecast AE normalized by the EuroCovidhub\_baseline AE. The values, which are highlighted in bold and orange color, correspond to the best performance. The lower part of the table reports for each method the number of countries for which its forecast is best, or in the top $2$, $3$ or $4$ best performing methods.}
    \label{tab:AE_w2}
\end{table}

\begin{table}[]
    \centering
\begin{tabular}{lrrrrrrr}
\toprule
       country &  EUHub-ens  &  IEM\_Health  & ILM& MUNI   &  RW  &  USC   &  SDSC\_ISG \\
\midrule 
       Austria &                      0.61 &                        1.44 &        1.03 & \textbf{\textcolor{orange}{0.45}} &                   1.10 &              0.53 &             0.69 \\
       Belgium &    \textbf{\textcolor{orange}{0.61}} &                        1.27 &        1.03 &           1.10 &                   1.33 &              1.01 &             1.09 \\
      Bulgaria &                      0.73 &                        1.49 &        1.46 & \textbf{\textcolor{orange}{0.50}} &                   1.12 &              0.92 &             1.02 \\
       Croatia &                      0.67 &                        1.22 &        1.15 & \textbf{\textcolor{orange}{0.53}} &                   0.89 &              0.92 &             0.88 \\
        Cyprus &                      0.84 &                        1.03 &        3.83 &  \textbf{\textcolor{orange}{0.69}} &                   1.21 &              1.10 &             0.81 \\
       Czechia &                      0.81 &                        0.83 &        1.41 &           0.67 &                   1.04 &              0.84 & \textbf{\textcolor{orange}{0.66}} \\
       Denmark &     \textbf{\textcolor{orange}{0.58}} &                        0.84 &        1.07 &           0.94 &                   1.20 &              1.06 &             0.63 \\
       Estonia &       \textbf{\textcolor{orange}{0.75}} &                        0.80 &        1.10 &           0.94 &                   1.08 &              1.02 &             0.98 \\
       Finland &      \textbf{\textcolor{orange}{0.75}} &                        0.76 &        1.07 &           1.04 &                   1.56 &              1.50 &             0.84 \\
        France &      \textbf{\textcolor{orange}{0.54}} &                        0.83 &        1.37 &           0.73 &                   1.20 &              1.01 &             0.59 \\
       Germany &                      0.57 &                        1.20 &        0.73 &           0.59 &                   0.85 &\textbf{\textcolor{orange}{0.55}} &             0.59 \\
        Greece &                      0.98 &                        1.53 &        1.70 & \textbf{\textcolor{orange}{0.73}} &                   1.63 &              1.49 &             1.14 \\
       Hungary &                      0.77 &                        1.51 &        1.13 & \textbf{\textcolor{orange}{0.54}} &                   1.18 &              0.68 &             0.58 \\
       Iceland &                      0.81 &                        1.52 &        1.94 &           0.55 &                   0.52 &              1.63 &             1.03 \\
       Ireland &                      1.37 &                        1.81 &        3.44 &\textbf{\textcolor{orange}{1.17}} &                   1.58 &              1.97 &             1.40 \\
         Italy &                      0.51 &                        0.85 &        0.90 &\textbf{\textcolor{orange}{0.44}} &                   1.23 &              1.15 &             0.61 \\
        Latvia &                      0.73 &                        1.07 &        1.00 &           0.98 &                   0.85 &\textbf{\textcolor{orange}{0.59}} &             0.60 \\
 Liechtenstein &                      0.78 &                        1.49 &        1.06 & \textbf{\textcolor{orange}{0.72}} &                   0.99 &              1.93 &             0.77 \\
     Lithuania &                      0.52 &                        0.97 &        1.36 &\textbf{\textcolor{orange}{0.52}} &                   1.06 &              0.98 &             0.91 \\
    Luxembourg &        \textbf{\textcolor{orange}{0.84}} &                        1.30 &        1.19 &           1.01 &                   1.42 &              2.28 &             1.41 \\
         Malta &                      1.25 &                        2.33 &       10.50 & \textbf{\textcolor{orange}{0.69}} &                   1.37 &              2.36 &             1.33 \\
   Netherlands &                      1.24 &                        1.00 &        3.10 &\textbf{\textcolor{orange}{0.80}} &                   1.20 &              0.93 &             1.02 \\
        Norway &                      0.69 &                        0.83 &        0.87 &           0.52 &                   1.11 &              1.31 &\textbf{\textcolor{orange}{0.48}} \\
        Poland &                      0.59 &                        1.13 &        1.04 &           0.31 &                   1.10 &              0.74 &             0.41 \\
      Portugal &      \textbf{\textcolor{orange}{0.69}} &                        0.83 &        1.14 &           1.11 &                   1.28 &              1.47 &             0.73 \\
       Romania &                      0.52 &                        1.21 &        0.79 &           0.39 &                   0.82 &              0.40 &             0.50 \\
      Slovakia &                      0.37 &                        0.46 & \textbf{\textcolor{orange}{0.36}} &           0.45 &                   0.88 &              0.61 &             0.49 \\
      Slovenia &                      0.94 &                        1.30 &        1.34 &\textbf{\textcolor{orange}{0.42}} &                   1.16 &              1.35 &             0.83 \\
         Spain &                      0.93 &                        1.19 &        2.34 &           1.09 &                   1.28 &              1.41 & \textbf{\textcolor{orange}{0.89}} \\
        Sweden &                      0.69 &                        1.06 &        0.70 &           1.31 &                   0.80 &              0.77 & \textbf{\textcolor{orange}{0.67}} \\
   Switzerland &         \textbf{\textcolor{orange}{0.55}} &                        1.08 &        0.98 &           0.80 &                   1.27 &              1.05 &             0.57 \\
\bottomrule 
    ranks best in   &           8 &                             0  &           1 &                 15 &                         1 &                    2 &               4 \\ 
    in top 2  &             19 &                                2  &           1 &                 19 &                         1 &                    6 &              14 \\ 
    in top 3  &              28 &                                5  &           4 &                 25 &                         1 &                    8 &              22 \\ 
    in top 4  &              30 &                                14  &          5 &                 28 &                        7 &                    11 &              29 \\ 
 \bottomrule
\end{tabular}
     \caption{Two week ahead forecast WIS normalized by the EuroCovidhub\_baseline WIS. The values, which are highlighted in bold and orange color, correspond to the best performance. The lower part of the table reports for each method the number of countries for which its forecast is best, or in the top $2$, $3$ or $4$ best performing methods.}
    \label{tab:WIS_w2} 
\end{table}

\end{document}